\documentclass[review]{elsarticle}

\usepackage[utf8]{inputenc}
\usepackage{todonotes}
\usepackage{float}
\usepackage[nolist]{acronym}
\usepackage{comment}
\usepackage{amsmath,amssymb,amsthm,amsfonts,enumitem}

\theoremstyle{plain}

\newtheorem*{definition*}{Definition}

\usepackage{scrextend}
\usepackage{verbatim}

\usepackage{array}
\newcolumntype{H}{>{\setbox0=\hbox\bgroup}c<{\egroup}@{}}
\newcolumntype{L}[1]{>{\raggedright\let\newline\\\arraybackslash\hspace{0pt}}m{#1}}
\newcolumntype{C}[1]{>{\centering\let\newline\\\arraybackslash\hspace{0pt}}m{#1}}
\newcolumntype{R}[1]{>{\raggedleft\let\newline\\\arraybackslash\hspace{0pt}}m{#1}}
\usepackage{longtable}
\usepackage{tabularx}
\usepackage{multirow}
\usepackage{pbox}
\usepackage{rotating}
\usepackage{makecell}
\usepackage{graphicx}
\usepackage{caption}
\usepackage{subcaption}
\usepackage{booktabs}
\usepackage[bookmarks, hidelinks]{hyperref}
\usepackage{xurl}
\usepackage{pdfpages}
\usepackage{bookmark}
\usepackage[section]{placeins}
\usepackage{bbm}

\usepackage[width=15cm, height=21.5cm]{geometry}
\usepackage{floatflt,epsfig, wrapfig}

\usepackage{natbib}
\setcitestyle{square}

\usepackage{lineno}
\modulolinenumbers[5]

\newif\ifdraft
\drafttrue
\ifdraft
\newcommand{\phnote}[1]{ {\textcolor{green} { ***Peter: #1 }}}
\newcommand{\jsnote}[1]{ {\textcolor{blue} { ***Johannes: #1 }}}
\newcommand{\trnote}[1]{ {\textcolor{red} { ***Timon: #1 }}}
\else
\newcommand{\phnote}[1]{}
\newcommand{\jsnote}[1]{}
\newcommand{\trnote}[1]{}
\fi

\makeatletter
\g@addto@macro\bfseries{\boldmath}
\makeatother

\renewcommand{\arraystretch}{1.25} 

\usepackage{pdfrender}

\usepackage{pifont}

\usepackage{listings, xcolor}

\definecolor{verylightgray}{rgb}{.97,.97,.97}

\usepackage{listings}
\definecolor{verylightgray}{rgb}{.97,.97,.97}
\definecolor{darkgreen}{rgb}{.0,.235,.0}
\lstdefinelanguage{Solidity}{
	keywords=[1]{anonymous, assembly, assert, balance, break, call, callcode, case, catch, class, constant, continue, constructor, contract, debugger, default, delegatecall, delete, do, else, emit, event, experimental, export, external, false, finally, function, gas, if, implements, import, indexed, instanceof, interface, internal, is, length, library, log0, log1, log2, log3, log4, memory, modifier, new, payable, pragma, protected, public, pure, push, require, return, returns, revert, selfdestruct, send, solidity, storage, struct, suicide, super, switch, then, this, throw, transfer, true, try, typeof, using, value, view, while, with, addmod, ecrecover, keccak256, mulmod, ripemd160, sha256, sha3}, 
	keywordstyle=[1]\color{blue}\bfseries,
	keywords=[2]{address, bool, byte, bytes, bytes1, bytes2, bytes3, bytes4, bytes5, bytes6, bytes7, bytes8, bytes9, bytes10, bytes11, bytes12, bytes13, bytes14, bytes15, bytes16, bytes17, bytes18, bytes19, bytes20, bytes21, bytes22, bytes23, bytes24, bytes25, bytes26, bytes27, bytes28, bytes29, bytes30, bytes31, bytes32, enum, int, int8, int16, int24, int32, int40, int48, int56, int64, int72, int80, int88, int96, int104, int112, int120, int128, int136, int144, int152, int160, int168, int176, int184, int192, int200, int208, int216, int224, int232, int240, int248, int256, mapping, string, uint, uint8, uint16, uint24, uint32, uint40, uint48, uint56, uint64, uint72, uint80, uint88, uint96, uint104, uint112, uint120, uint128, uint136, uint144, uint152, uint160, uint168, uint176, uint184, uint192, uint200, uint208, uint216, uint224, uint232, uint240, uint248, uint256, void, ether, finney, szabo, wei, days, hours, minutes, seconds, weeks, years},	
	keywordstyle=[2]\color{teal}\bfseries,
	keywords=[3]{block, blockhash, coinbase, difficulty, gaslimit, number, timestamp, msg, data, gas, sender, sig, value, now, tx, gasprice, origin},	
	keywordstyle=[3]\color{violet}\bfseries,
	identifierstyle=\color{black},
	sensitive=false,
	comment=[l]{//},
	morecomment=[s]{/*}{*/},
	commentstyle=\color{gray}\ttfamily,
	stringstyle=\color{darkgreen}\ttfamily,
	morestring=[b]',
	morestring=[b]"
}
\lstdefinelanguage{circom}{
	keywords=[1]{assert, do, else, false, function, if, length, public, true, while, for}, 
	keywordstyle=[1]\color{blue}\bfseries,
	keywords=[2]{var, signal},	
	keywordstyle=[2]\color{teal}\bfseries,
	keywords=[3]{include, template, input, output, private, main, component},	
	keywordstyle=[3]\color{violet}\bfseries,
	identifierstyle=\color{black},
	sensitive=false,
	comment=[l]{//},
	morecomment=[s]{/*}{*/},
	commentstyle=\color{gray}\ttfamily,
	stringstyle=\color{darkgreen}\ttfamily,
	morestring=[b]',
	morestring=[b]"
}
\journal{Computer Networks}

\title{Fairness, Integrity, and Privacy in a Scalable Blockchain-based Federated Learning System}
\author{\texorpdfstring{\mbox{Timon Rückel\,$^{\mathrm{a},\ast}$}}}
\author{\texorpdfstring{\mbox{Johannes Sedlmeir\,{$^{\mathrm{b},\mathrm{c}}$}}}}
\author{\texorpdfstring{\mbox{Peter Hofmann\,{$^{\mathrm{b},\mathrm{c}}$}}}}
\address{
    $^\mathrm{a}$
    University of Bayreuth, Bayreuth, Germany \\
    $^\mathrm{b}$
    FIM Research Center, University of Bayreuth, Bayreuth, Germany \\
    $^\mathrm{c}$
    Project Group Business \& Information Systems Engineering of the Fraunhofer FIT, Bayreuth, Germany \\~\\
    $^\ast$
    Corresponding author: \href{mailto:timonrueckel@hotmail.de}{timonrueckel@hotmail.de}
    \\~\\~\\~\\~\\
    \large{\textup{This is the accepted version of an article with the same name, published in the Special Issue ``Federated Learning and Blockchain Supported Smart Networking in Beyond 5G (B5G) Wireless Communication'' in Computer Networks.}}
}

\date{\today}

\begin{document}

\begin{frontmatter}
\begin{abstract}
    Federated machine learning (FL) allows to collectively train models on sensitive data as only the clients' models and not their training data need to be shared. However, despite the attention that research on FL has drawn, the concept still lacks broad adoption in practice. One of the key reasons is the great challenge to implement FL systems that simultaneously achieve fairness, integrity, and privacy preservation for all participating clients. To contribute to solving this issue, our paper suggests a FL system that incorporates blockchain technology, local differential privacy, and zero-knowledge proofs. Our implementation of a proof-of-concept with multiple linear regression illustrates that these state-of-the-art technologies can be combined to a FL system that aligns economic incentives, trust, and confidentiality requirements in a scalable and transparent system.
\end{abstract}
\end{frontmatter}

	\clearpage
\pagenumbering{gobble} 
\textbf{Highlights}
\begin{itemize}
    \item Fairness, integrity, and privacy are important requirements for federated learning.
    \item It is challenging to achieve all of these requirements at the same time.
    \item With blockchain, differential privacy, and zero-knowledge proofs we can satisfy them.
    \item We provide a proof of concept for the case of multiple linear regressions.
    \item Our evaluation indicates that the architecture is practical and scalable.
\end{itemize}
\vspace{3em}
\begin{figure}[H]
    \centering
    \includegraphics[page=1, width=\linewidth, trim=0cm 0cm 0cm 0cm, clip]{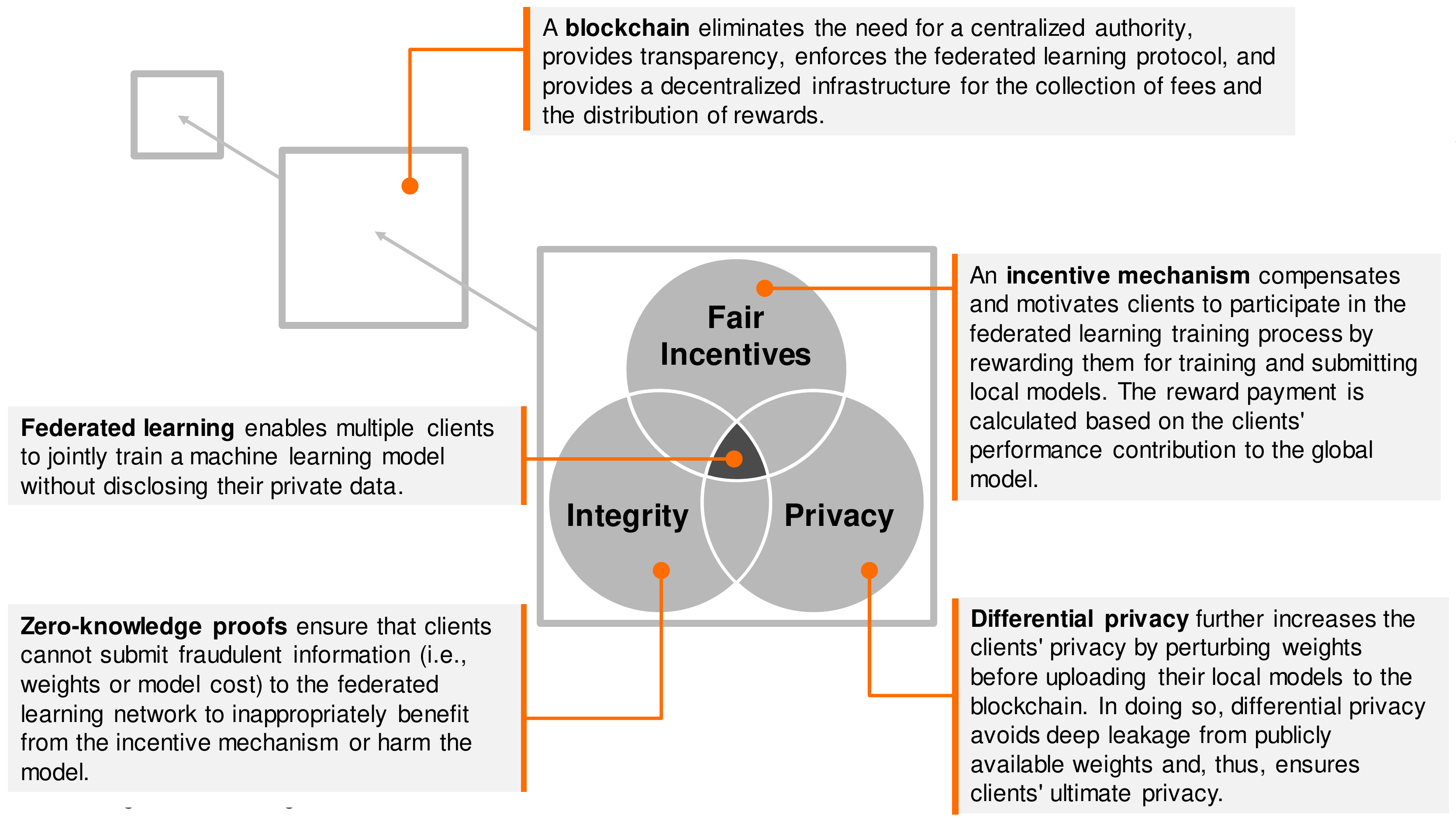}
    \label{fig:highlights}
\end{figure}
\clearpage

\acrodefplural{SHARK}[SHARKs]{succinct hybrid arguments of knowledge}
\acrodefplural{SNARK}[SNARKs]{succinct non-interactive arguments of knowledge}
\acrodefplural{STARK}[STARKs]{scalable transparent arguments of knowledge}

\begin{acronym}[SHARK]
\setlength{\itemsep}{-0.1cm}

    \acro{DL}{distributed machine learning}
    \acro{DP}[LDP]{local differential privacy}
    \acro{_DP}[DP]{differential privacy}
    \acro{EVM}[EVM]{Ethereum virtual machine}
    \acro{FedAvg}[FedAvg]{federated averaging}
    \acro{FL}[FL]{federated machine learning}
    \acro{FHE}[FHE]{fully homomorphic encryption}
    \acro{GDPR}[GDPR]{general data protection regulation}
    \acro{HFL}[HFL]{horizontal federated learning}
    \acro{IID}[i.i.d.]{independent and identically distributed}
    \acro{IoT}[IoT]{internet of things}
    \acrodef{LOO}[LOO]{leave-one-out}
    \acro{LR}[LR]{linear regression}
    \acro{ML}[ML]{machine learning}
    \acro{MPC}[MPC]{multiparty computation}
    \acro{PKI}[PKI]{public key infrastructure}
    \acro{RSS}[RSS]{residual sum of squares}
    \acro{SNARK}[SNARK]{succinct non-interactive argument of knowledge}
    \acroplural{SNARK}[SNARKs]{succinct non-interactive arguments of knowledge}
    \acro{STARK}[STARK]{scalable transparent argument of knowledge}
    \acro{ZKP}[ZKP]{zero-knowledge proof}
\end{acronym}



\clearpage
\pagenumbering{arabic}  
    \setcounter{page}{1}
  
\section{Introduction}
\label{sec:introduction}

    The application of \ac{ML} promises far-reaching potentials across industries~\citep{Iansiti.2020}. \ac{ML} has already proven successful in many areas, such as web search or recommender systems in e-commerce, in which a lot of high-quality data exists~\citep{LeCun.2015}. While researchers address \ac{ML}'s growing demand for compute power and use of data with, e.g., distributed \ac{ML} approaches where multiple computing nodes share their resources~\citep{Galakatos2018DistributedML, Kairouz_2019_Advances, Verbraeken2019Survey} and quality issues with data processing, access to data is not only a technical issue. Both traditional \ac{ML} and distributed \ac{ML} approaches assume that their training data is centralized by nature, preventing the applicability of \ac{ML} approaches to domains in which data is sensitive and distributed at the same time. To avoid that \ac{ML} approaches must rely on data to which only a centralized organization or individual has full access, \ac{FL} can aggregate the less sensitive \ac{ML} models that were independently and locally trained by individual clients~\citep{Li_2020_Federated,Yin2021flsurvey}. Consequently, \ac{FL} can enable the use of \ac{ML} applications in domains with strong privacy requirements and contribute to solving the challenge of limited access to sensitive data without invading participating clients' privacy~\citep{Larson_2020_Ethics,yang2019federated}. For instance, researchers at Google aimed to improve next word predictions for mobile devices based on private user data, i.e., the words users are typing~\citep{DBLP:journals/corr/abs-1811-03604}. In the case of autonomous driving, \ac{FL} could reduce the data transmission overhead in vehicular networks while still respecting privacy requirements~\citep{elbir2020federated}. These examples demonstrate \ac{FL}'s capability to avoid the obligation to centralize data. Thus, \ac{FL} approaches improve functionality or even enable new value creation scenarios for \ac{ML}.
    
    Despite these promising applications and developments, \ac{FL} in practice has not yet encountered broad adoption~\citep{Aledhari_2020_Federated}. This can be traced to a variety of design requirements that have not been met simultaneously so far.
    For example, \ac{FL} systems require, amongst others, privacy guarantees exceeding \ac{FL}'s privacy by design~\citep{Kaissis2020Secure} as well as high degrees of fairness and integrity~\citep{Aledhari_2020_Federated,Yang_2019_Federated}.
    Compared to centralized \ac{ML}, \ac{FL} already ensures a certain level of privacy for participating clients~\citep{Aledhari_2020_Federated}. However, even when replacing centralized data with clients' model updates in a \ac{FL} system, these public model updates can still leak insights on private client data~\citep{Kaissis2020Secure,Hitaj2017Deep, Melis2019Exploiting, Nasr2019Comprehensive, Phong2018Privacy, Zhu_2020_Deep}. While research has addressed this issue with different approaches, there are tradeoffs in terms of performance and integrity, as plausibility checks of the contributed models cannot be made any more when the individual model updates are obfuscated. Second, \ac{FL} systems can be subject to malicious client attacks that try to harm the global model performance by submitting model updates that have been trained on data sets unequal to the client's actual data (or even generated on purpose to harm the quality of the aggregate global model). So-called data-poisoning attacks can reduce model performance by up to 90\,\%~\citep{yang2019federated,Aledhari_2020_Federated}.
    Third, the above application examples assume that users contribute their data, computation, and communication resources unconditionally. However, scaling \ac{FL} to broad adoption in practice requires fair and transparent incentive mechanisms to appropriately remunerate clients for their contribution to the global model performance~\citep{Kairouz_2019_Advances,Aledhari_2020_Federated,Yang_2019_Federated,Faltings2017Game,Li_2020_review}. \ac{FL} can be subject to free-riding attacks in which malicious clients fraudulently benefit from the incentive mechanism by submitting model updates that are not based on the respective client’s private data~\citep{Aledhari_2020_Federated, Yang_2019_Federated}. Moreover, in a conventional \ac{FL} setting, clients are forced to trust the central entity to remunerate all clients fairly without being able to check whether this central entity acts truthfully or maliciously.
    
    Satisfying the described privacy, integrity, and fairness requirements in a \ac{FL} system whilst still being scalable requires an interdisciplinary discourse that bundles up a combination of technologies within a \ac{FL} system~\citep{Kairouz_2019_Advances, Li_2020_Federated}.
    Even though a vast amount of research in various disciplines already exists, there are, to the best of our knowledge, no systems that jointly deliver privacy, integrity, and fair incentives.
    Thus, we explore the following research question:\\
    \begin{addmargin}[2em]{2em}
        \textit{How can a \ac{FL} system achieve fairness, integrity, and privacy whilst still being practical and scalable?}\\
    \end{addmargin}
    
    To answer this research question, we propose a \ac{FL} system that levers blockchain technology, \acf{DP}, and \acp{ZKP}. We thus integrate these emerging technologies to provide a novel and smart \ac{FL}-based architecture with the following properties:
    \begin{itemize}
        \item \textbf{Fair incentives}: Our proposed \ac{FL} architecture measures the individual contribution to the global model performance per client based on the client's actual parameters (i.e., without the \ac{DP}-noise) and incentivizes each client accordingly. By building our \ac{FL} system based on blockchain, a smart contract enforces the transparent and verifiable distribution of incentives~\citep{Kurtulmus2018Trustless, Mugunthan2020Blockflow}.
        \item \textbf{Integrity}: Non-interactive \acp{ZKP} enable clients to validate that fellow clients have truthfully trained their submitted model updates based on private data that they committed to earlier, potentially including a proof of provenance (e.g., from a certified sensor). In doing so, these fellow clients do not have to reveal any of their private data, yet we can guarantee that they do the training and evaluation for their incentive truthfully. Further, we build our \ac{FL} system based on blockchain. In the resulting decentralized setting, there is no trust in a central authority needed regarding censorship and the correct aggregation as well as the availability of the global model. Research has already pointed out that a blockchain can be a suitable replacement for intermediaries in a collaborative process and help achieve standardized communication between participants~\citep{bokolo2021collaboration}. Besides, the blockchain-based design ensures neutrality amongst all clients, the immutability of transactions, and the full transparency of the architecture for all clients.
        \item \textbf{Privacy}: To make sure that clients' model updates cannot leak information on patterns within their private data, we leverage \ac{DP} to perturb each clients' model update with Laplacian noise.
    \end{itemize}
  
    We discuss and instantiate our architecture for multiple \ac{LR} as \ac{FL} model. For the implementation, we use \acp{SNARK} implemented in \href{https://github.com/iden3/circom}{circom} and \href{https://github.com/iden3/snarkjs}{snarkjs} as well as smart contracts deployed on an \ac{EVM} implemented in \href{https://soliditylang.org/}{Solidity}. 
    Through implementing and testing the system, we demonstrate that a realization of our architecture can achieve reasonable performance and scalability. Moreover, we gain valuable insights into how the combination of \ac{FL}, \ac{DP}, \acp{ZKP}, and blockchain can be applied to more sophisticated \ac{ML} models beyond \ac{FL}.
    By proposing our \ac{FL} system that integrates several emerging technologies in a novel way, we contribute a solution that demonstrates that fairness, integrity, and privacy requirements can be solved in practical settings and thus also improve the real-world applicability of existing \ac{FL} approaches. As we pointed out that using \ac{FL} approaches alone is not practical for some application scenarios, we contribute to overcoming relevant hurdles towards applicability. The developed \ac{FL} system is scalable, ensures integrity, and attracts clients to participate through fair incentives as well as data privacy guarantees.
    
    We structure this paper as follows: In Section~\ref{sec:background}, we briefly present the technical building blocks of our \ac{FL} system (i.e., \ac{FL}, \ac{DP}, \acp{ZKP}, and blockchain technology) and discuss related work on the design of \ac{FL} systems. Afterward, we present the architecture and implementation of our \ac{FL}-system in Section~\ref{sec:architecture} and evaluate it in Section~\ref{sec:evaluation}. Finally, we discuss our results, describe our research's limitations, and outline future research opportunities in Section~\ref{sec:conclusion}.

\section{Foundations}
\label{sec:background}

\subsection{Federated Learning}
    
    \Acl{FL} describes the concept of training local \ac{ML} models on distributed and private client data without transferring the data beyond the client's reach. After training the local parameters, clients in the \ac{FL} system submit model updates derived from their locally trained parameters to a  server that aggregates all local model updates to a global model~\citep{Kairouz_2019_Advances}. The types of potential clients are diverse and range from organizations and mobile devices equipped with sensors to autonomous vehicles. Research has recently applied \ac{FL} systems in several domains including health, the \ac{IoT}, vehicular networks, finance, sales, or smart homes~\citep{Li_2020_Federated, yang2019federated}.
    By bringing the computation to the data, \ac{FL} improves clients' privacy~\citep{Li_2020_Federated}. Besides, \ac{FL} can increase the efficiency of existing infrastructure and devices by avoiding the transfer of large data sets to a central server and by utilizing the computational power of edge devices like smartphones or wearables~\citep{Li_2020_review}. In doing so, \ac{FL} is typically associated with the following optimization problem~\citep{Li_2020_Federated}: 
    \begin{align}
        \min\limits_{w_g} F(w_g) = \min\limits_{w_g} \left( \sum\limits_{i=1}^{|I|} b_i F_i(w_g) \right) ~\text{,}
    \end{align}
    where $|I|$ is the total number of clients, $b_i \geq 0$, $\sum_{|I|} b_i = 1$ denotes the client's relative impact, and $F_i(w)$ is the local objective function. The global, aggregated weight $w_g$ is mostly derived from the local updates $w_i$ using \ac{FedAvg}, an aggregation scheme that computes a weighted average of the local weights $w_i$~\citep{Nilsson_2018_Performance}. When training on \ac{IID} data, \ac{FedAvg} achieves similar results to centralized learning~\citep{Nilsson_2018_Performance}.

\subsection{Local Differential Privacy}
    
    Differential privacy (DP) has been developed by \citet{Dwork2006Calibrating} to allow for analyzing sensitive and private data in a secure way:
    Consider a trusted central authority that holds a data set containing sensitive client information. The key idea of DP is to develop a query function on the sensitive data set that returns the true answer plus random noise following a carefully chosen distribution~\citep{Dwork2014Algorithmic}. For example, in a study that asks participants to report a certain personal property, participants report their binary answer by tossing a coin: They respond truthfully if tails and if heads, the participants toss the coin again and report ``yes" if heads and ``no" if tails. In this simple example, the participants' privacy stems from the plausible deniability of any reported value~\citep{Dwork2014Algorithmic}.
    
    The driving force for this privacy guarantee is randomization since the guarantee must hold regardless of all present or even future sources of background information (e.g., from the internet or newspapers). Achieving this requires understanding the input and output space of randomized algorithms. Formally, a randomized algorithm $\mathcal{M}$ with domain $A$ and (discrete) range $B$ is associated with a mapping from $A$ to $\Delta (B)$, the probability simplex over $B$:
    \begin{align}
        \Delta (B) = \Bigg\{ x \in \mathbb{R}^{|B|}: x_i \geq 0 ~\mathrm{for~all}~ i ~\mathrm{and}~ \sum\limits_{i=1}^{|B|} x_i=1 \Bigg\}.
    \end{align}
    Then, on input $a \in A$ and given that the probability space is over the coin flips of $\mathcal{M}$, the randomized algorithm $\mathcal{M}$ outputs $\mathcal{M}(a) = b$ with probability $\left( \mathcal{M}(a) \right)_b$ for every $b \in B$~\citep{Dwork2014Algorithmic}.
    Before defining DP, note that DP aims to sanitize a query function such that the presence or absence of an individual in the analyzed data set cannot be determined by just observing the output of the query. Now consider two data sets $D$ and $D'$, that are either equal or differ only in the presence or absence of one individual, as histograms. Moreover, a histogram over a universe $\mathcal{X}$ is an object in $\mathbb{N}^{\mathcal{|X|}}$. Then, both data sets $D, D' \in \mathbb{N}^{\mathcal{|X|}}$ are called adjacent if for the $\ell_1$-norm, $\lVert D - D' \rVert_1 \leq 1$ holds. Eventually, \citet{Dwork2006Calibrating} define DP as follows:
    \begin{definition*}
    \label{def:DP}
        A randomized algorithm $\mathcal{M}$ with domain $\mathbb{N}^{\mathcal{|X|}}$ and range $\mathcal{R}$ satisfies $\epsilon$-differential privacy if for every adjacent data sets $D, D' \in \mathbb{N}^{\mathcal{|X|}}$ and any subset $\mathcal{S} \subseteq \mathcal{R}$ we have
        \begin{align}
            \mathrm{Pr}[ \mathcal{M}(D) \in \mathcal{S} ] \leq e^\epsilon~\mathrm{Pr}[ \mathcal{M}(D') \in \mathcal{S} ],
        \end{align}
        where $\epsilon>0$ is a privacy parameter.
    \end{definition*}
    A common way to achieve DP for a numeric function $f$ is to add noise following a Laplacian distribution:
    \begin{align}
    \label{equation:LapDistribution}
        \mathcal{L}(x~|~0,\lambda) &= \frac{1}{2\lambda} \exp{\left( -\frac{|x|}{\lambda} \right).}
    \end{align}
    When a user wants to learn $f(\mathbf{x}) = \sum_i x_i$ with $\mathbf{x}\in \{0,1\}$, i.e., the total number of 1's in the data set, $\epsilon$-DP can be achieved by adding Laplacian noise~\citep{Dwork2014Algorithmic, Dwork2010Differential}:
    \begin{align}
    \label{equation:noisysum}
        \Tilde{f}(\mathbf{x}) = \sum_i x_i + q, \qquad \mathrm{where}~ q \sim \mathcal{L} \left(0, \frac{\Delta f}{\epsilon} \right).
    \end{align}
    and $\Delta f$ denotes the sensitivity of $f$, i.e., the maximum difference of $f$ on data sets that differ in only one element~\citep{Dwork2010Differential}.
    There are many practical algorithms for providing DP; and in general, they have many characteristics beyond $\varepsilon$ that have an impact on privacy or utility~\citep{garrido2021get}. One special case of DP is \acf{DP}, where the confusion (i.e., random perturbation) is performed locally by clients and not by a central authority~\citep{Dwork2004Privacy}. By doing so, the central authority cannot infer or access the actual client data. According to Definition \ref{def:DP}, anyone accessing $\mathcal{S}$ cannot distinguish whether the true data set is $D$ or $D'$ with confidence (controlled by $\epsilon$ -- the lower $\epsilon$, the higher the privacy and the lower the accuracy and vice versa). Thus, \ac{DP} ensures plausible deniability for the clients~\citep{Nguyen2016Collecting}.

\subsection{Verifiable Computation and Zero-Knowledge Proofs}
\label{subsec:ZKP}

The notion of \acp{ZKP} was first introduced by~\citet{goldwasser1989knowledge}. \acp{ZKP} are a special form of protocols between a so-called \textit{prover} and a \textit{verifier} in which the prover wants to convince the verifier that she/he knows some value with a specific property (more formally, an element of a language). \acp{ZKP} have the additional property that the prover learns \textit{nothing} beyond this statement. 
The most important properties beyond zero-knowledgeness that we build upon in this paper are \textit{completeness} (an honest prover will convince the verifier with high probability if the statement is correct) and \textit{soundness} (any and in particular a malicious prover will convince the verifier of a false statement only with a small probability). By replacing the verifier with a random oracle such as a hash function, a large class of interactive proofs can be transferred to non-interactive proofs (using the so-called Fiat-Shamir heuristic~\citep{fiat1986prove}). As opposed to interactive \acp{ZKP}, prover and verifier do not have to interact with each other in non-interactive \acp{ZKP}. Notably, there are \emph{succinct} \ac{ZKP}, which means that both the proof size as well as the computational complexity of the proof verification is considerably smaller than the complexity of checking the statement by conducting the original computation~\citep{bitansky2013succinct}.

Since the introduction of \ac{ZKP}, there has been a lot of research on them, but practical implementations or even applications remained rare before the beginning of the 2010s. However, starting with~\citet{groth2006perfect}, a period of rapid development of \ac{ZKP} towards practical implementations delivered significant performance improvements, e.g., in~\cite{ben2013snarks}. In recent years, different flavors of non-interactive \acp{ZKP} have emerged, for example, Bulletproofs~\citep{bunz2018bulletproofs}, \acp{SNARK}~\citep{gennaro2013quadratic}, \acp{STARK}~\citep{ben2019scalable}, or hybrid constructions. While they differ in scaling properties and cryptographic assumptions, they all allow creating proofs for the correct execution of a program without displaying all inputs, outputs or intermediate steps. Often, Merkle proofs for the inputs are revealed to commit the prover to the usage of unknown but fixed variables. Domain-specific languages such as \href{https://github.com/iden3/circom}{Circom} or \href{https://github.com/starkware-libs/cairo-lang}{Cairo} allow compiling programs into arithmetic circuits. From these, polynomials are constructed, which in turn can be translated into proving and verification programs through libraries such as \href{https://github.com/scipr-lab/libsnark}{libsnark} or \href{https://github.com/iden3/snarkjs}{snarkjs}. As such generic tools for \ac{ZKP} have significantly matured over the last years, they have increasingly been used in first applications; often associated with blockchains and distributed ledgers, where due to redundant execution, cheap verification without revealing sensitive data is important~\citep{ben2019scalable}. However, the generality to which the correctness of computations can be proved with these frameworks is often limited to prime field operations, complemented by libraries that provide, e.g., circuits for basic cryptographic and arithmetic operations such as hash functions, signature schemes, and comparators.
\subsection{Blockchains and Distributed Ledger Technology}
\label{subsec:DLT}

Blockchain and, more general, distributed ledger technology builds upon peer-to-peer networks in which all data is replicated, shared, and distributed across multiple servers (`nodes')~\citep{butijn2020blockchains}. In blockchains, an append-only structure connects batches of transactions (`blocks') linearly through hash-pointers (`chain') and thus achieves decentralized yet synchronized data management. A so-called consensus mechanism that typically combines cryptographic techniques with economic or social incentives allows deciding which blocks to append as well as the order of transactions within a block~\citep{xiao2020consensus}.
If a majority of the network in a specific metric like hash rate (proof of work), the share of cryptocurrency (proof of stake), or the number of accounts in a permissioned network (voting-based consensus mechanisms) is honest, this guarantees the correct execution of simple payments and programming logic (``smart contracts'') and the practical immutability of the ledger~\citep{xiao2020consensus}. The confidence that a majority of the network behaves as intended without the need to rely on the honesty of a distinguished entity is often referred to as digital trust~\citep{nofer2017blockchain}. Consequently, blockchains allow avoiding dependencies on one or a few distinct entities on digital platforms~\citep{alt2020electronic,fridgen2018cross}. The literature distinguishes between permissionless blockchains (such as those used in cryptocurrencies) where anyone can participate and permissioned blockchains where participation is limited, e.g., to an industry consortium or the public sector~\citep{wust2018you}.

Since the release of the Bitcoin whitepaper~\citep{nakamoto2008peer}, blockchain technology has been used in various applications, e.g., cryptocurrencies, decentralized finance with derivatives and non-fungible tokens, or industry applications.
One early and popular permissionless blockchain that supports a Turing-complete programming language for smart contracts is Ethereum. It provides a decentralized virtual machine environment, namely the \acf{EVM}, for executing smart contracts. Ethereum smart contracts are usually implemented in \href{https://soliditylang.org/}{Solidity}, a high-level programming language with syntax similar to \href{https://www.javascript.com/}{JavaScript}~\citep{Wohrer2018Smart}. Two special properties of \href{https://soliditylang.org/}{Solidity} are the lack of non-deterministic libraries (that would otherwise conflict with the necessarily deterministic design of a blockchain that first orders and then executes transactions) and that the complexity of execution has a price, counted in so-called \emph{gas}. This avoids not only infinite loops but facilitates fair competition for the limited capacity.

However, as blockchains and distributed ledgers exhibit \textit{redundant} storage and computation, they suffer from major challenges. While a high energy consumption is often presumed, in fact, only proof of work blockchains are problematic in this regard~\citep{sedlmeir2020energy}. Two other issues that arise directly from replicated transaction storage and execution are considerably more fundamental: Scalability~\citep{gudgeon2020sok} and privacy~\citep{zhang2019security}. Yet, there are innovative approaches to mitigate these challenges. For scalability, countermeasures range from restricting participation and demanding high computational power, storage, and bandwidth from the participating nodes to sharding and off-chain computations. Off-chain computations are also good for privacy, but lead to the challenge of verification in a system with malicious participants. On the other hand, methods for privacy are technologies like \ac{DP}, \ac{FHE}, or \ac{MPC}~\citep{munilla2021revealing}. \acp{ZKP} can be regarded as a special case of verifiable \ac{MPC} where only one participant contributes private data but the result is verified by all blockchain nodes. Yet, \acp{ZKP} are arguably significantly closer to broad adaption and have been leveraged by many blockchain projects so far, starting with \href{https://z.cash/}{Z-Cash} and now also covering many scalability and privacy projects, many of which are implemented on Ethereum (e.g., \href{https://tornado.cash/}{Tornado-Cash}, \href{https://loopring.org/#/}{Loopring}, \href{https://aztec.network/}{Aztec}, \href{https://www.starkdex.io/}{StarkDEX}).
\subsection{Related Work}
\label{sec:related_work}
    
Since the term \acl{FL} was introduced by \citet{McMahan2017Communication}, research has focused on improving, amongst others, performance, privacy, integrity, and incentive-mechanisms~\citep{Aledhari_2020_Federated,Yang_2019_Federated}.
To improve the clients' privacy beyond the level that \ac{FL} inherently offers, research came up with three main strategies, namely homomorphic encryption, \ac{MPC}, and DP, which aim to prevent public model updates from leaking private client information. Due to its low complexity and strong privacy guarantee~\citep{Kairouz_2019_Advances}, DP is widely used~\citep{Li_2020_Federated}, even though deploying DP in \ac{ML} leads to a trade-off between maximizing privacy (i.e., adding noise with high variance) and maximizing accuracy. In practice, instead of uploading the actual weights, clients can add DP noise to their weights. For example~\citep{Lu2020Differentially} developed a \ac{FL} system for vehicular networks that combines \ac{DP} with gradient descent to avoid attacks that leak private information from publicly available \ac{ML} model updates. Instead of perturbing model parameters, \ac{DP} noise can also be added to the training data~\citep{Yin2021flsurvey}. However, this approach cannot provide privacy protection since it is not sufficient to make any single record unnoticeable~\citep{domingo2021limits}.
    
Besides, many works have adopted game-theoretic approaches to motivate clients' participation in a \ac{FL} system and ensure fairness amongst them. As an example, \citet{Khan_2020_Federated} implemented a Stackelberg game to incentivize clients for contributing to training a model and, at the same time, maximize the model's performance. Since clients must trust the central authority to incentivize all clients fairly in a traditional \ac{FL} setting and to provide a decentralized incentive layer for data sharing in general~\citep{jaiman2021user}, researchers have suggested using blockchains and smart contracts for model aggregation (e.g., \citep{Rmanan2020Baffle}) and client remuneration (e.g., \citep{Toyoda2019Mechanism} or \citep{Kang_2019_Incentive}).
\citet{jin2021fl} propose a \ac{FL} architecture that leverages a blockchain for transparency and incentivizing clients and combines it with homomorphic encryption in the aggregation smart contract to prevent leakage from clients' contributions.
However, in all these frameworks, offering incentives to clients also puts the system's integrity at risk as malicious clients may try to fraudulently benefit from the incentive mechanism, e.g., through free-riding attacks~\citep{Aledhari_2020_Federated, Yang_2019_Federated}.
    
The transparency of smart contracts could help achieve integrity by allowing clients to recalculate the weights that fellow clients submitted. However, even when ignoring the corresponding privacy and scalability challenges, such an approach would lead to a ``verifier's dilemma''~\citep{Luu201Demystifying}, where clients weigh up between accepting the costs of recalculation or trusting other clients. \Acp{ZKP} could offer an efficient solution to the ``verifier's dilemma'' in \ac{FL} and, thus, pave the way to achieving fairness, integrity, and privacy at the same time. Despite \acp{ZKP}' potential for \ac{FL} systems and their increasing adoption (especially in the blockchain domain), it remains an open question how \acp{ZKP} can be used in the context of \ac{FL}~\citep{Kairouz_2019_Advances}. 
\citet{Wu2018DIZK} have implemented a \ac{SNARK} that proves the correctness of \ac{LR} parameters by recalculating them. In the case of \ac{LR}, this can be done only using matrix multiplications. However, this approach includes rounding a matrix inverse and, hence, requires further measures to ensure full tamper protection (as we will show in detail in Section~\ref{sec:arch_conarch_parameters}). \citet{Feng2021ZEN} introduced a toolchain to produce verifiable and privacy-preserving \acp{SNARK} that prove correct inference in classification and recognition tasks by taking an existing neural network as input. Also \citet{Zhang2020Zero} as well as \citet{weng2021mystique} implemented \acp{ZKP} that allow verifying whether a particular prediction by a trained \ac{ML} model has been computed truthfully without providing any information about the \ac{ML} model itself. Even though also \citet{weng2021mystique}'s work implements a \ac{ZKP} merely for \ac{ML} model inference, they improved \acp{ZKP}' efficiency to prove the correctness of matrix multiplications and \acp{ZKP}' application to floating point arithmetic, which are both essential ingredients for training \ac{ML} models.
    
Despite the advancement that these works generate for combining \acp{ZKP} and \ac{ML}, it remains, to the best of our knowledge, still unclear how recent progress in \acp{ZKP}, blockchain technology, and DP can be combined as a technology stack that achieves full privacy, integrity, and fairness in \ac{FL} systems.

\section{Architecture and Implementation}
\label{sec:architecture}

We organize Section~\ref{sec:architecture} as follows: In Section \ref{sec:arch_conceptual_architecture} we explain how our suggested \ac{FL} system is built up formally. To do so, we divide the \ac{FL} process into four different sections, namely  \nameref{sec:arch_conarch_parameters} (Section \ref{sec:arch_conarch_parameters}), \nameref{sec:arch_conarch_aggregation} (Section \ref{sec:arch_conarch_aggregation}), \nameref{sec:arch_conarch_cost} (Section \ref{sec:arch_conarch_cost}), and \nameref{sec:arch_conarch_incentives} (Section \ref{sec:arch_conarch_incentives}). Table~\ref{tab:notation} introduces the notation that we use throughout this section.
\begin{table}[!htb]
    \centering
    \resizebox{0.6\linewidth}{!}{
    \renewcommand{\arraystretch}{0.8}
    \setlength{\aboverulesep}{0pt}
    \setlength{\belowrulesep}{0pt}
    \begin{tabular}{|c|l|}\toprule
         \textbf{Symbol} & \textbf{Explanation} \\\midrule
         $I$ & Set of indices, $i\in\{1,\ldots,\lvert I\rvert\}$ \\
         $u_i$ & Client $i$, $i\in I$ \\
         $k$ & Number of features of the linear regression \\
         $n$ & Sample size per client of the linear regression (set globally) \\
         $d$ & Accuracy of rounded inputs \\
         $\mathbf{X}_i$ & $n \times (k+1)$ Matrix of input data including all independent variables \\
         $Y_i$ & $n$-dimensional vector of input data containing all dependent variables \\
         $\mathbf{Z}_i$ & Approximate inverse of $(\mathbf{X}_i^\intercal \mathbf{X}_i)^{-1}$ \\
         $\mathbf{D}_i$ & Private data set of $u_i$ -- an $n \times (k+2)$ matrix: $(X_{i,0}, X_{i,1}\ldots X_{i,k}~Y_i)$ \\
         $\mathbf{D}_{\textrm{test}}$ & Public test data set \\
         $rt^\mathbf{D}_i$ & Merkle tree root of $\mathbf{D}_i$ \\ 
         $rt^{\mathbf{D}_{\mathrm{test}}}_i$ & Merkle tree root of $\mathbf{D}_{\textrm{test}}$ \\
         $l_{\mathrm{train}}$ & Depth of $\mathbf{D}_i$'s Merkle tree (number of levels) \\
         $l_{\mathrm{test}}$ & Depth of $\mathbf{D}^{\mathrm{test}}_i$'s Merkle tree (number of levels) \\
         $\varepsilon_{\mu}$ & Upper bound for $\mu\:\cdot\:n$ \\
         $\varepsilon_{\sigma}$ & Upper bound for $\sigma\:\cdot\:n$ \\
         $\varepsilon_{\mathrm{inverse}}$ & Upper bound for $\lVert(\mathbf{X}^\intercal \mathbf{X}) \mathbf{Z} - \mathbbm{1}\rVert$ \\
         $\varepsilon_w$ & Upper bound for $\lVert w - \tilde{w} \rVert$ \\
         $\varepsilon_{w^{\prime}}$ & Upper bound for $\lVert w^{\prime} - \tilde{w}^\prime \rVert$ \\
         $\vartheta_{\mathbf{X}^\intercal Y}$ & Upper bound for $\lVert \mathbf{X}^\intercal Y \rVert$ \\
         $\vartheta_{\mathbf{Z}}$ & Upper bound for $\lVert \mathbf{Z} \rVert$ \\
         $M$ & Model (corresponds to respective weight vector) \\
         $w_{i}$ & Weight vector of $u_i$, corresponding to $M_{i}$ \\
         $w_{i}^\prime$ & Weight vector including \ac{DP} noise \\
         $\tilde{w}_{i}$ & Recalculated weight vector \\
         $\tilde{w}_{i}^\prime$ & Recalculated weight vector including \ac{DP} noise \\
         $w_{\mathrm{g}}$ & Weight vector of global model $M_{\mathrm{g}}$ \\
         $\varepsilon_{\mathrm{LR}}$ & Error term of \ac{LR} \\
         $L$ & Vector including Laplacian \ac{DP} noise \\
         $d_{\mathcal{L}}$ & Accuracy of the Laplacian \ac{DP} noise \\
         $q$ & \ac{DP} noise \\
         $h_j$ & Hash serving as random noise for $q$ \\
         $hash\_alg$ & Hash algorithm, i.e., $0$ for `MiMC' or $1$ for `Poseidon' \\
         $C$ & Vector of $u_i$'s cost: $(c_1, \dots, c_{|I|})$ \\
         $V$ & Vector of $u_i$'s incentive payment: $(v_1, \dots, v_{|I|})$ \\
         $B$ & Admission fee payable upon joining the system \\
         $\pi(a,b)$ & \ac{ZKP} for statement $a$ with witness $b$ \\
         Gen($\pi$) & Construct the \ac{ZKP} $\pi$ (with the proving key) \\
         Ver($\pi$) & Verify \ac{ZKP} $\pi$ (with the verification key) \\\midrule
    \end{tabular}
    }
    \caption{Notation used in this paper.}
    \label{tab:notation}
\end{table}

\subsection{Conceptual Architecture}
\label{sec:arch_conceptual_architecture}

\subsubsection{Compute and Prove Model Weight}
\label{sec:arch_conarch_parameters}

    At the start, participating clients $u_i$ can register at the smart contract \verb|Clients|. To do so, they must commit to the Merkle root $rt^\mathbf{D}_i$ of their private data set $\mathbf{D}_i = (\mathbf{X}_i~Y_i) = (X_{i,0}~X_{i,1}~\ldots~X_{i,k}~Y_i)\in\mathbb{R}^{n \times (k+2)}\sim\mathbb{R}^{n(k+2)}$ with 
    \begin{align*}
        \mathbf{X}_{i} = 
        \begin{pmatrix}
            x_{i,0,1} & x_{i,1,1} & \cdots & x_{i,k,1} \\
            x_{i,0,2} & x_{i,1,2} & \cdots & x_{i,k,2} \\
            \vdots  & \vdots  & \ddots & \vdots  \\
            x_{i,0,n} & x_{i,1,n} & \cdots & x_{i,k,n} 
        \end{pmatrix}
        =
        \begin{pmatrix}
            1 & x_{i,1,1} & \cdots & x_{i,k,1} \\
            1 & x_{i,1,2} & \cdots & x_{i,k,2} \\
            \vdots  & \vdots  & \ddots & \vdots  \\
            1 & x_{i,1,n} & \cdots & x_{i,k,n} 
        \end{pmatrix}
        \in \mathbb{R}^{n \times (k+1)}
    \end{align*}
    and 
    \begin{align*}
        Y_i = \left(y_{i,1}, \ldots, y_{i,n} \right)^{\intercal} \in \mathbb{R}^n \textrm{.}
    \end{align*}
    Prior to computing $rt^\mathbf{D}_i$ and training their \ac{LR} weights, clients must normalize their input data $\mathbf{D}_i$, such that every $X_{i,1},\ldots,X_{i,k},Y_i$ has expectation value \mbox{$\mu=0$} and standard deviation \mbox{$\sigma=1$} (in line with, e.g., \citet{Witten2017Data}). Subsequently, clients train their \ac{LR} weight \mbox{$w_i = (\beta_{i,0}~\beta_{i,1} \ldots \beta_{i,k})^{\intercal}\in\mathbb{R}^k$} by computing
    \begin{align}
    \label{equation:LinReg_params}
        w_i = (\mathbf{X}_i^\intercal \mathbf{X}_i)^{-1} ~ \mathbf{X}_i^\intercal ~ Y_i.
    \end{align}
    Equation \eqref{equation:LinReg_params} optimizes
    \begin{align}
    \label{equation:LinReg_RSS}
        \min (\text{RSS}) = \min\limits_{w\in\mathbb{R}^k} \left( (Y_i - \mathbf{X}_i w_i)^{\intercal} (Y_i - \mathbf{X}_i w_i) \right)=\lVert\varepsilon_{\mathrm{LR}, i}\rVert_2
    \end{align}
    where $Y_i = \mathbf{X}_i w_i + \varepsilon_{\mathrm{LR}, i}$ and $\varepsilon_{\mathrm{LR}, i}$ denotes the \ac{LR}'s error term. 
    Constructing the \ac{ZKP} $\pi^w$ that ensures that $u_i$ computed $w_i$ truthfully (i.e., truly computed the $w_i$ that minimizes the \ac{RSS} using \eqref{equation:LinReg_RSS} based on $u_i$'s originally committed data) requires adapting the protocol to \href{https://github.com/iden3/circom}{circom}'s capabilities that are restricted to prime field operations and, thus, to non-negative integer inputs. To do so, we decide to generally round and subsequently scale all (public and private) \ac{ZKP} inputs $\mathbf{A^{\lvert S \rvert \times \lvert T \rvert}} = (a_{s,t})_{s=1,\ldots,\lvert S \rvert; ~ t=1,\ldots,\lvert T \rvert}$ as in general we cannot assume that $a_{s,t} \in \mathbb{N}_0$. For example, we convert the input $a_{s,t} = 2.5347725$ to $2.53477 \cdot 10^5 = 253477 = \tilde{a}_{s,t}$ in the cae of $d = 5$. Further, to input only positive integers, we introduce a matrix $\mathbf{Sign(A)}$ whose elements indicate the signs of $\mathbf{A}$'s coefficients: $\left(\mathbf{{Sign}(A)}\right)_{s,t}=\mathrm{sign}(a_{s,t})$ where 
    \begin{align*}
        \mathrm{sign}: \mathbb{Z}\to\{-1,1\}, \quad x\mapsto 
            \begin{cases}
                0 & \text{if $x \geq 0$} \\
                1 & \text{if $x < 0$}
            \end{cases}.
    \end{align*}
    Then we can write the \ac{ZKP} inputs as 
    \begin{align}
    \label{equation:input_conversion}
        \tilde{\mathbf{A}} = 10^{-d}\,\mathbf{{Sign}(A)}\circ \mathbf{\tilde{A}^+}
        \qquad \text{where} \quad \tilde{a}^+_{s,t}\in \mathbb{N}_0 \text{.} 
    \end{align}
    Moreover, \mbox{$\lVert \mathbf{\tilde A} - \mathbf{A}\rVert\leq c\cdot 10^{-d}$}, where $c$ depends on the choice of matrix norm. For example. if we use the maximum norm, $c$ can be $0.5$ times the number of columns of $A$.
    
    After this transformation, $u_i$ can compute the \ac{ZKP}. However, as constructing proofs is computationally expensive, designing the \ac{ZKP} generation requires specific consideration and optimizations. For better readability, we will refrain from using client indices $i$ in the following description.
    First, to make sure that the client's input data is standardized, we introduce an upper bound for the mean $\mu$. Note that since \href{https://github.com/iden3/circom}{circom} does not allow for non-integer divisions, we apply the upper bound controlling $\mu$ to the sum of all elements in $Y$ and $X_1, \ldots X_k$: 
    \begin{align}
    \label{equation:epsilon_mu}
    \begin{split}
        \mu_{X_j} \cdot n = \sum\limits_{l=1,\ldots n} x_{j,l} & \leq \varepsilon_{\mu} \\
        \mu_Y \cdot n = \sum\limits_{l=1,\ldots n} y_{l} & \leq \varepsilon_{\mu}
    \end{split} ~\quad~\forall~j\in\{1,\ldots,k\}.
    \end{align}
    Similarly, we regulate the data set's variance $\sigma^2$ by applying an upper bound to the squared sums:
    \begin{align}
    \label{equation:epsilon_sigma}
    \begin{split}
        \sigma_{X_j}^2 \cdot n = \sum\limits_{l=1,\ldots n}\left(x_{j,l}-\mu\right)^2 & \leq \varepsilon_{\sigma}\\
        \sigma_Y^2 \cdot n = \sum\limits_{l=1,\ldots n}\left(y_{l}-\mu\right)^2 & \leq \varepsilon_{\sigma}
    \end{split} ~\quad~\forall~j\in\{1,\ldots,k\}. 
    \end{align}
    Hence, $\varepsilon_{\mu}$ and $\varepsilon_{\sigma}$ must be set with respect to $n$.
    
    Recalling \eqref{equation:LinReg_params}, our initial attempt was to compute the inverse $(\mathbf{X}^\intercal \mathbf{X})^{-1}$ within a circuit of the respective \ac{ZKP} by introducing two integers, a denominator and a numerator, per input value. This procedure would allow us to calculate $(\mathbf{X}^\intercal \mathbf{X})^{-1}$ in the circuit (as divisions leading to non-integer values can be replaced by multiplications) without rounding errors (given the input values are precise). However, the approach turned out to have one significant shortcoming: Computing $(\mathbf{X}^\intercal \mathbf{X})^{-1}$ within a circuit using Gaussian elimination quickly results in overflow (i.e., a numerator or denominator can quickly get larger than the size of the prime field). Finding common denominators often requires multiplying all denominators repeatedly. For example, already computing the inverse of $(\mathbf{X}^\intercal \mathbf{X})^{5 \times 5}$ matrices with $d=5$ can cause an overflow (in general, this will depend on the choice of $d$ and $k$, but $2^k\cdot d$ will probably be larger than $78$ in many practical applications). On the other hand, reducing fractions or rounding (using range proofs) before overflow would add further complexity in the Gaussian elimination algorithm. Besides, more complex operations demanded by more sophisticated \ac{ML} algorithms would require to hand-craft a fast numerical solver, so the method does not generalize well. Eventually, this approach would hardly scale to more complicated optimization algorithms.

    As matrix inversion inside the circuit or inverting within \href{https://github.com/iden3/circom}{circom}'s prime field is expensive, we decided to follow the approach of~\citet{Wu2018DIZK} to solve the problem by letting clients calculate and provide an inverse $\mathbf{Z} \approx (\mathbf{X}^\intercal \mathbf{X})^{-1}$ \emph{outside the circuit} through a standard solver such as typical matrix inversion tools in \href{https://www.javascript.com/}{Javascript}, \href{https://www.python.org/}{Python}, or \href{https://de.mathworks.com/products/matlab.html}{MATLAB}. We then give this inverse as private input to the circuit (again complying with the above-described conversion to non-negative integer entries). However, due to the finite precision of the matrix inversion, we cannot assume that equality holds. Thus, we apply a slightly different method and only prove the proximity of the calculated inverse to the true inverse. To do so, we check the approximation error of the provided numerical inverse using the following range constraint:
    \begin{align}
    \label{equation:epsilon_inverse}
        \lVert(\mathbf{X}^\intercal \mathbf{X}) \mathbf{Z} - (\mathbf{X}^\intercal \mathbf{X})(\mathbf{X}^\intercal \mathbf{X})^{-1}\rVert = \lVert(\mathbf{X}^\intercal \mathbf{X}) \mathbf{Z} - \mathbbm{1}\rVert \leq \varepsilon_{\mathrm{inverse}} < 1.
    \end{align}
    As matrix norm, the maximum norm is a convenient choice, i.e., the inequality can be checked index-wise (respecting $k$). Through this bound, we can both control for the effect that the replacement of $(\mathbf{X}^\intercal \mathbf{X})^{-1}$ with its approximation $\mathbf{Z}$ has and ensure that the system remains protected against malicious client attacks. 
    To do so, we estimate an upper bound for the approximation effects on the distance from $w$ to $\tilde{w}$, with the latter being calculated based on $\mathbf{Z}$, via
    \begin{align}
    \label{equation:epsilon_w}
    \begin{split}
        \lVert w - \tilde{w} \rVert
        &\leq \lVert (\mathbf{X}^\intercal \mathbf{X})^{-1} \mathbf{X}^\intercal Y - \mathbf{Z} \mathbf{X}^\intercal Y \rVert\\
        &\leq \lVert(\mathbf{X}^\intercal \mathbf{X})^{-1}-\mathbf{Z}\rVert \cdot \lVert \mathbf{X}^\intercal Y \rVert \leq \varepsilon_w ,
    \end{split}
    \end{align}
    where we use the submultiplicativity of $\lVert\,\cdot\,\rVert$. Even though all entries of $\mathbf{X}^\intercal$ and $Y$ are normalized, i.e., $X_{1},\ldots,X_{k},Y$ have expectation value $\mu=0$ and the standard deviation $\sigma=1$, $\lVert \mathbf{X}^\intercal Y \rVert$ might still yield large entries. Thus, to control the error $\lVert w - w^\prime \rVert$ between the true weights and the client's approximative result, we desire upper bounds for both $\lVert (\mathbf{X}^\intercal \mathbf{X})^{-1} - \mathbf{Z} \rVert$ and $\lVert \mathbf{X}^\intercal Y \rVert$.
    
    First, we introduce a simple range proof to ensure that $\lVert \mathbf{X}^\intercal Y \rVert$ is bounded:
    \begin{align}
        \lVert \mathbf{X}^\intercal Y \rVert \leq \vartheta_{\mathbf{X}^\intercal Y} 
        \quad \mathrm{where}\quad\vartheta_{\mathbf{X}^\intercal Y} \equiv \vartheta_{\mathbf{X}^\intercal Y}\left(k,d\right) \text{.}
    \end{align}
    In general, the probability that $\lVert \mathbf{X}^\intercal Y \rVert$ is large is very small as the normal distribution decays fast; if by coincidence a single value is big, $d$ might need to be adjusted to increase the accuracy of rounded inputs.

    Second, we recall that the true inverse can essentially be replaced by the approximative inverse as long as the approximate inverse is bounded~\citep{Newman1974How}. More precisely, given any matrix norm $\lVert\,\cdot\,\rVert$, a square nonsingular matrix $\mathbf{A}$ and its approximate inverse $\mathbf{Z}$ such that \mbox{$\lVert\mathbf{A} \mathbf{Z} - \mathbbm{1}\rVert \leq \varepsilon_{\mathrm{range}} < 1$}, the bound
    \begin{align}
    \label{eq:newman_ineq}
        \lVert \mathbf{A}^{-1}-\mathbf{Z} \rVert \leq \frac{\lVert \mathbf{Z} \rVert \times \lVert\mathbf{A} \mathbf{Z} - \mathbbm{1} \rVert}{1 - \lVert\mathbf{A}\mathbf{Z} - \mathbbm{1} \rVert}
    \end{align}
    holds. However, it is not clear that $\mathbf{X}^\intercal \mathbf{X}$ is non-singular, and checking this inside the circuit, e.g., through computing the determinant, would become expensive already for moderate $k$. Fortunately, we can relax the assumption by reconsidering the argument in the derivation of the result in~\citet{Newman1974How}: Provided the candidate for the approximate inverse satisfies $\lVert \mathbf{R}\rVert <1$ for $\mathbf{R}:= \mathbf{A}\mathbf{Z} - \mathbbm{1}$, then the Neumann-series $\mathbbm{1} + \mathbf{R} + \mathbf{R}^2 + \ldots$ converges. Consequently, using the argument in the paper,
    \begin{align}
        \mathbf{A}\mathbf{Z}\cdot(\mathbbm{1} + \mathbf{R} + \mathbf{R}^2 + \ldots)=(\mathbbm{1}-\mathbf{R})\cdot(\mathbbm{1} + \mathbf{R} + \mathbf{R}^2 + \ldots)=\mathbbm{1}.
    \end{align}
    It follows that $\mathbf{Z}(\mathbbm{1} + \mathbf{R} + \mathbf{R}^2 + \ldots)$ is a right-inverse of the quadratic matrix $\mathbf{A}$ and hence its unique (left- and right-) inverse, i.e., $\mathbf{A}$ is non-singular. In particular, we learn that if $\mathbf{A}$ was non-singular, we could not find an approximate inverse that satisfies~\eqref{equation:epsilon_inverse}. Applied to $\mathbf{A}:=\mathbf{X}^\intercal \mathbf{X}$, our circuit already implicitly guarantees that $\mathbf{X}^\intercal \mathbf{X}$ is invertible through checking~\eqref{equation:epsilon_inverse}, and~\eqref{eq:newman_ineq} yields
    \begin{align}
    \label{eq:error_bound_inv}
        \lVert (\mathbf{X}^\intercal \mathbf{X})^{-1}-\mathbf{Z} \rVert \leq \frac {\lVert \mathbf{Z} \rVert \times \lVert(\mathbf{X}^\intercal \mathbf{X}) \mathbf{Z} - \mathbbm{1}\rVert} {1 - \lVert(\mathbf{X}^\intercal \mathbf{X}) \mathbf{Z} - \mathbbm{1}\rVert} \leq \vartheta_{\mathbf{Z}} \cdot \varepsilon_{\mathrm{inverse}} ~\text{,}
    \end{align}
    where $\quad\rVert \mathbf{Z} \lVert ~\leq~ \vartheta_{\mathbf{Z}}$ and $\vartheta_{\mathbf{Z}} \equiv \vartheta_{\mathbf{Z}}\left(k,d\right)$.
    
    Before uploading $w$ to the \ac{EVM} storage, clients must perturb their weight by adding Laplacian \ac{DP}-noise (recall~\eqref{equation:LapDistribution} and~\eqref{equation:noisysum}) to avoid deep leakage~\citep{Zhu_2020_Deep}: 
    \begin{align}
        \label{equation:DP-noise}
        w^{\prime} &= w + q, \qquad && q \sim \mathcal{L}(0, \lambda), ~q \in \mathbb{R}^k \\
        \label{equation:DP-noise_lambda}
        \lambda &= \frac{\Delta}{\epsilon} = \frac{1}{\epsilon} \left( \max \beta_{i,j} - \min \beta_{i,j} \right) \qquad &&i \in \{1,\dots, \lvert I \rvert\}, ~j \in \{0,\dots, k\}.
    \end{align}
    Note that every entry in $q$ is drawn separately, following $\mathcal{L}(0, \lambda)$. Moreover, since we perturb client weights that will be aggregated, we must set the \ac{DP} security parameter $\epsilon$ globally. As no fellow $\beta_{i,j}$ will be available to any client before uploading the client's own $w$, we must set an estimation for $\Delta$ globally. Recalling that all $w$ are computed based on normalized data, we know that $\beta_0,\dots,\beta_k \in [\text{-}1;1]$ holds. Therefore, we set $\Delta = 2 \cdot 10^d$ as a proxy. Moreover, to generate $\pi^w$, we must ensure that we can reproduce every entry in $q$ while keeping its choice random. Achieving this requires a twofold approach.
    
    First, since \href{https://github.com/iden3/circom}{circom} does not have tools to compute a Laplace-distributed random variable directly, we discretize $\mathcal{L}(0,\lambda)$, introducing a variable $d_{\mathcal{L}} = 10^x$ with $x \in \mathbb{N}$ that defines the interval of a discrete distribution $\mathcal{DL}(0,\lambda)$ of $\mathcal{L}(0,\lambda)$ (i.e., $\mathcal{DL}$'s accuracy). We then construct a vector $L \in \mathbb{R}^{d_{\mathcal{L}}-1}$ whose entries are given by the inverse cumulative distribution function of $\mathcal{DL}(0,\lambda)$:
    \begin{align}
    \label{equation:DP_L}
        L(p) := -\lambda~\text{sgn}\left(p-\frac{1}{2} \right)~\text{ln}\left(1-2\Big|p-\frac{1}{2}\Big|\right), \quad p = \{1,\dots, d_{\mathcal{L}}-1\}.
    \end{align}
    The finite nature of machines leads to some form of discretization of distributions, which makes DP mechanisms generally vulnerable to attacks~\citep{Mironov2012Significance}. Therefore, there is a research stream dedicated to exploring the effects of discretization on \ac{DP} (e.g., \citeauthor{Balcer2019Differential} (\citeyear{Balcer2019Differential}), \citeauthor{Canonne2021Discrete} (\citeyear{Canonne2021Discrete})). However, in our system, the parameters are currently set in a way that the limitation through $d_{\mathcal{L}}$ will be stricter than that of the precision typically achieved in computer programs.

\begin{figure}[!htb]
        \begin{subfigure}{.5\textwidth}
            \centering
            \includegraphics[page=1, width=\linewidth, trim=10cm 6cm 10cm 6cm, clip]{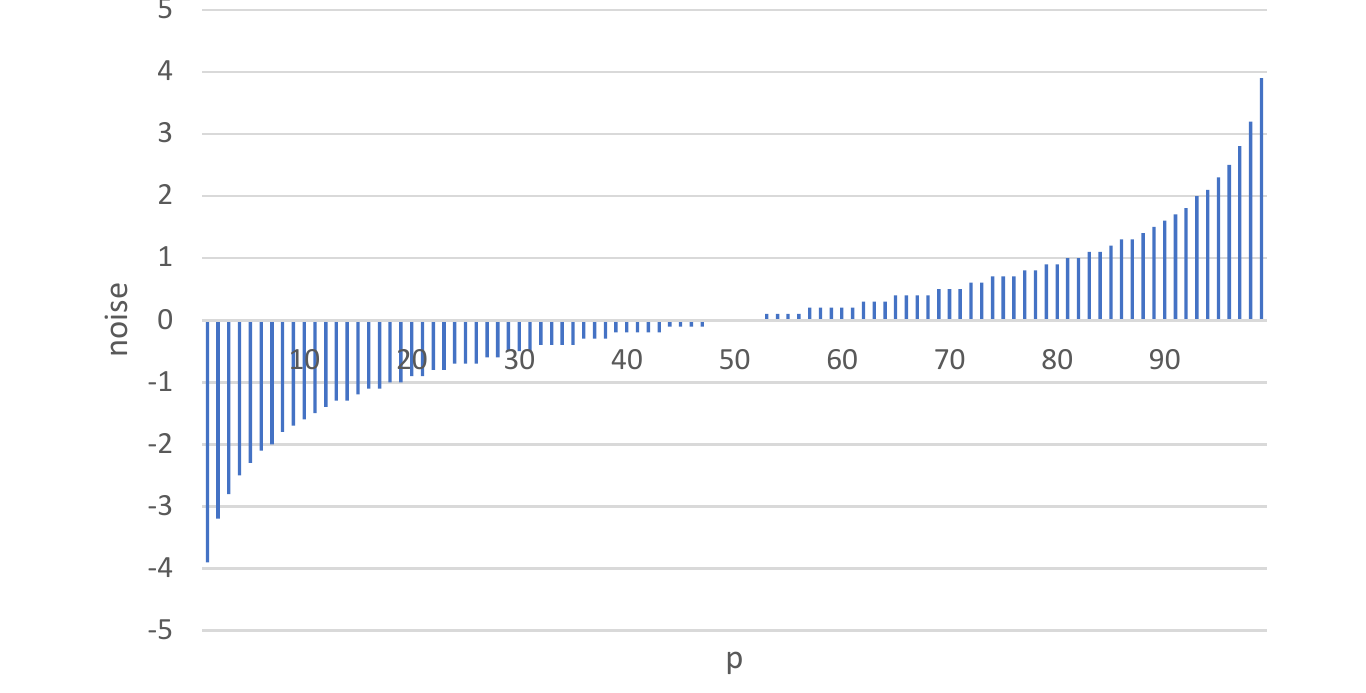}
            \caption{Value of the noise depending on $p$.}
            \label{plot:discrete_noise_QQ}
        \end{subfigure}
        \begin{subfigure}{.5\textwidth}
            \centering
            \includegraphics[page=2, width=\linewidth, trim=10cm 6cm 10cm 6cm, clip]{Figures/Lagrange-plots.pdf}
            \caption{Histogram of the discretized Laplacian noise.}
            \label{plot:discrete_noise_histogram}
        \end{subfigure}
        \caption{Derivation of the discretized Laplacian noise.}
    \end{figure}

    Second, we draw the underlying randomness $p$ from the random oracle $h_j$ that results from hashing a solid source of entropy like the current block hash of the blockchain and the entry of $Y$ that corresponds to the $\beta_j$ that is being perturbed:
    \begin{align}
    \label{equation:DP_hash_j}
        h_j = H\left(\text{current block hash}~|~y_j \right), \quad j=\{0,\dots,k\}.
    \end{align}
    So, for example, when perturbing $\beta'_0 = \beta_0 + q_0$, the corresponding entry of $Y$ is $y_0$. The resulting $h_j$ is a 256-bit number. We derive every $p$ by computing
    \begin{align}
    \label{equation:DP_p}
        p = h_j ~\text{mod}~ d_{\mathcal{L}}, \quad j=\{0,\dots,k\} \text{.}
    \end{align}
    The noise that results from a particular value of $p$, as well as its distribution (which is close to the true Laplace distribution with $\lambda=1$, is illustrated in Figures~\ref{plot:discrete_noise_QQ}~and~\ref{plot:discrete_noise_histogram}).
    Essentially, we generate verifiable randomness (similar as used, e.g., in Algorand for electing block producers) deterministically in the circuit by combining entropy from the blockchain and from the training data that was the client previously committed to. This approach ensures that clients cannot influence their \ac{DP}-noise $q$ whilst fellow clients cannot reproduce the $q$ (which would allow them to leak the `unperturbed' $w$) and verify that the noise was produced purely at random: As the client could not predict the entropy taken from the blockchain at the time of committing to $D_i$, and consequently not try different values for $Y$ that yield the desired noise. Additionally, by inputting~$L$, $\pi_w$'s circuits can verify $p$.
    Now the clients are ready to generate their \acp{ZKP} that allows for verifying the validity of the clients' local computations of $w^{\prime}$. Given the proposed architecture, the \ac{ZKP} $\pi^w(a,b)$ with statement $a$ and witness $b$ that proves that $u$ has truthfully calculated $w^{\prime}$ has the following characteristic:
    \begin{align}
        \pi^w \left( \left( w^{\prime}, rt^{\mathbf{D}}, \varepsilon_{\mu}, \varepsilon_{\sigma}, \varepsilon_{w^{\prime}}, \varepsilon_{\mathrm{inverse}}, \vartheta_{\mathbf{X}^\intercal Y}, \vartheta_{\mathbf{Z}} \right), 
        \left( \mathbf{D}, \mathbf{Z} \right) \right).
    \end{align}
    Besides the above-described range proofs and computation checks, $\pi^w$ ensures a globally set range $\varepsilon_{w^{\prime}}$ for the submitted $w^{\prime}$:
    \begin{align*}
        \lVert w^{\prime} - \tilde{w}^{\prime} \rVert \leq \varepsilon_{w^{\prime}}.
    \end{align*}
    Please refer to Section \ref{sec:arch_implementation} for a detailed description of how we implement $\pi^w$.
    
    Using a smart contract and through snarkjs' \verb|export solidityverifier| command, clients can, subsequently to compiling the circuit and generating the proof, upload their $w^{\prime}$ to the \ac{EVM}. The smart contract verifies the proof and, if valid, uploads the client's $w^{\prime}$.

\subsubsection{Model Aggregation}
\label{sec:arch_conarch_aggregation}
    
    For every successfully uploaded $w^{\prime}_i$, the smart contract \verb|Clients| updates $w_{\mathrm{g}}$ by aggregating all weights that have been submitted to the smart contract so far using FedAvg. For a globally constant sample size $n$ in a \ac{LR} setting, the respective aggregation formula equals
    \begin{align}
    \begin{split}
        w_{\mathrm{g}} = \frac{\sum_{i=1}^{\lvert I \rvert} w_i}{\lvert I \rvert} \approx \frac{\sum_{i=1}^{\lvert I \rvert} w^{\prime}_i}{\lvert I \rvert}= w^{\prime}_{\mathrm{g}} \text{,}
    \end{split}
    \end{align}
    depending on $\epsilon$ and $\lvert I \rvert$. Heuristically, the noise is \ac{IID} with expectation value $0$ and finite variance $2\lambda^2$, so by the law of large numbers, the probability that $w_{\mathrm{g}}$ differs significantly from $w^{\prime}_{\mathrm{g}}$ is small (in fact, the deviation is approximately normally distributed with standard deviation $\tfrac{2\lambda^2}{\sqrt{|I|}}$.

\subsubsection{Compute and Prove Model Cost}
\label{sec:arch_conarch_cost}
    
    After submitting and proving their \ac{DP}-weights $w^{\prime}$, clients test their unperturbed $w$ on a global and public data set \mbox{$\mathbf{D}_{\textrm{test}} = (\mathbf{X}_{\mathrm{test}}~Y_{\mathrm{test}}) = (X_{\mathrm{test},0} ~X_{\mathrm{test},1} ~\ldots~X_{\mathrm{test},k}~Y_{\mathrm{test}})\in\mathbb{R}^{n_{\mathrm{test}} \times (k+2)}$} with $X_{\mathrm{test}} = (1, ..., 1)^{\intercal}$,  distributed via a public channel (e.g., a cloud provider). Clients have to prove that they computed their model cost $c$ truthfully based on $\mathbf{D}_{\textrm{test}}$ and using their $w$ by constructing a \ac{ZKP} $\pi^c$ with the following characteristic.
    \begin{align}
    \label{eq:pi^c}
        \pi^c \left( \left( c, rt^{\mathbf{D}}, rt^{\mathbf{D}_{\mathrm{test}}}, \varepsilon_w \right), \left( \mathbf{D}, \mathbf{Z}, w \right)  \right).
    \end{align}
    $\pi^c$ ensures that clients used the private input $w$ (i.e., their true and unperturbed weight) to compute $c$ by first reproducing $\tilde{w}$ in a circuit of the \ac{ZKP} and performing the range proof (recall \eqref{equation:epsilon_w})
    \begin{align*}
        \lVert w - \tilde{w} \rVert \leq \varepsilon_w \quad \text{.}
    \end{align*}
    Note that the SC \verb|Clients| requires that clients have uploaded a valid $w$ prior to submitting and proving $c$, such that the standardization of $rt^{\mathbf{D}}$ as well as the effects of approximating $\lVert \mathbf{X}^{\intercal} Y \rVert$ through $\lVert \mathbf{Z} \rVert$ are already controlled by $\pi^w$.
    Then, as $w$ is a private input to $\pi^c$, we can compute $\tilde{c}$ in a circuit of $\pi^c$ and make sure that
    \begin{align}
    \label{equation:cost}
        \tilde{c} =
        \big( Y_{\mathrm{test}} - \hat{Y}_{\mathrm{test}} \big)^{\intercal} \big( Y_{\mathrm{test}} - \hat{Y}_{\mathrm{test}} \big)
        = c
    \end{align}
    with $\hat{Y}_{\mathrm{test}} = \mathbf{X}_{\mathrm{test}}\,\cdot\,w$ holds.
    
    Second, by including $rt^{\mathbf{D}_{\mathrm{test}}}$ as a statement, $\pi^c$ can compute the respective Merkle tree on the test data set $\mathbf{D}^{\mathrm{test}}$ and require that the resulting root $\tilde{rt}^{\mathbf{D}_{\mathrm{test}}}$ equals $rt^{\mathbf{D}_{\mathrm{test}}}$, with the latter being stored in the \ac{EVM}. Since $\mathbf{D}_{\mathrm{test}}$ is publicly available, all clients can easily check its standardization, such that checking $\tilde{rt}^{\mathbf{D}_{\mathrm{test}}} = rt^{\mathbf{D}_{\mathrm{test}}}$ is sufficient. This approach allows the system to guarantee truthful submissions of $c$ without having to store $\mathbf{D}_{\mathrm{test}}$ onchain, which would, depending on $k$ and $n_{\mathrm{test}}$
    be costly and challenging. A detailed implementation of $\pi^c$ is again included in Section \ref{sec:arch_implementation}.

\subsubsection{Compute Incentives}
\label{sec:arch_conarch_incentives}
    
    To translate $C = (c_1, \dots, c_{|I|})^{\intercal}$ transparently into each client's incentive payment $v_i$, we choose an efficient approach that can be computed on-chain by the \verb|Clients| smart contract. First, we normalize all submitted and validated $c_i$'s to a vector $C_{\mathrm{norm}} \in \mathbb{R}^{|I|}$ that has expectation value \mbox{$\mu=0$} and the standard deviation \mbox{$\sigma=1$}. We multiply all entries in $C_{\mathrm{norm}}$ with $(-1)$ as lower $c_i$ implies higher model performance and, subsequently, replace negative entries with $0$ (i.e., ensure that all $c_{\mathrm{norm,}i}$ that are higher than the mean of all $c_i$ are not incentivized). Next, we scale the data such that $\sum\limits_{i=0}^{|I|} c_{\mathrm{norm},i} = 1$. Upon entry, clients have to pay an admission fee $B$, such that we can distribute the incentive payments $V = (v_1, \dots, v_{|I|})^{\intercal} \in \mathbb{R}^{+}$ as follows: 
    \begin{align}
        V = B \cdot |I_{\mathrm{valid}}| \cdot C_{\mathrm{norm}}
    \end{align}
    where $|I_{\mathrm{valid}}|$ is the number of clients with valid $w^{\prime}_i$ and $c_i$.
    The payments are distributed by the function \verb|reward_clients()| which can only be executed by the client who has initially deployed the  \verb|Clients| smart contract. Moreover, by using \href{https://soliditylang.org/}{Solidity}'s \verb|mapping|, every client (i.e., address) can only register once as the data would be overwritten when joining multiple times and a new \verb|clientID| would be set. We summarize the system outlined above in Table \ref{tab:architecture}.

\begin{table}[!htb]
    \centering
    \scriptsize
    \renewcommand{\arraystretch}{1.5}
    \begin{tabularx}{\textwidth}{|c|p{7cm}|X|}\toprule
        \textbf{Step} & \textbf{Client \texorpdfstring{$u_{i}$}{ui}} & \textbf{Smart Contract} \\\midrule
        \midrule
        \multicolumn{3}{|c|}{Training phase}\\
        \midrule
        1 & \textbf{Join system:} Join the system by fetching the model $M_i$ (i.e., number of features $k$ and the sample size $n$), hashing $u_i$'s database, writing $rt^\mathbf{D}_i$ into storage, and paying the admission fee $B$ & \\
        
        2 & \textbf{Train model:} Train $M_i$ by computing $w_i$ & \\
        
        3 & \textbf{Compute and upload \ac{DP} weight $w^{\prime}_i$:} Compute and upload differentially private (\ac{DP}) weight $w^{\prime}_i$ & \\
        
        4 & \textbf{Compute and submit weight proof $\pi^w_i$:} \newline  Gen$\left( \pi^w_i \right) \space \rightarrow \space \pi^w_i$ and upload $\pi^w_i$. $w^{\prime}_i$ will only be uploaded into storage if $\pi^w_i$ is valid & \\

        5 & & \textbf{Verify all \ac{DP} weights:} \newline Ver$\left(\pi^w_i\right)~\forall~i\in\{1,\ldots,\lvert I\rvert\}$. Write every $w^{\prime}_i$ into storage if $\pi^w_i$ has been validated \\
        
        \midrule
        \multicolumn{3}{|c|}{Testing phase}\\
        \midrule
        
        6 & & \textbf{Global model aggregation:} Aggregate all $w'_i$ to $w^{\prime}_{\mathrm{g}} \approx w_{\mathrm{g}}$ and write $w^{\prime}_{\mathrm{g}}$ into storage\\
        
        7 & \textbf{Compute $u_i$'s cost $\mathrm{c}_i$:} Fetch the public test data set and evaluate own model performance (\ac{RSS}) using $u_i$'s true weight $w_i$. Submit own accuracy as $u_i$'s cost $c_i$ & \\
        
        8 & \textbf{Compute and upload cost proof $\pi^c_i$:} \newline Gen$ \left( \pi^c_i \right) \space \rightarrow \space \pi^c_i$ and upload $\pi^c_i$. $c_i$ will only be uploaded into storage if $\pi^c_i$ is valid & \\
        
        9 & & \textbf{Verify all cost proofs:\newline} Ver$\left(\pi^c_i\right)~\forall~i\in\{1,\ldots,\lvert I\rvert\}$ \\
        
        10 & & \textbf{Compute incentives:} Compute and distribute incentive payments $V$ according to $C$ \\
        \midrule
    \end{tabularx}
    \caption{Steps of our protocol for private, fair, and honest \ac{FL}.}
    \label{tab:architecture}
\end{table}

\subsection{Implementation}
\label{sec:arch_implementation}

In this section, we describe the implementation of our architecture. Please note that, following \href{https://github.com/iden3/circom}{circom} terminology and in contrast to Section~\ref{sec:arch_conceptual_architecture}, we will now refer to single `circuits' that are callable with input and output signals as `templates', whereas we refer to a set of `templates' that make up a \ac{ZKP} as `circuits'. `Signals' (\verb|signal|) are the basis for defining constraints (using \verb|===|) and can only be assigned a value once (using \verb|<--| or \verb|-->|). Both \verb|<==| and \verb|==>| declare a constraint and assign a signal's value at the same time. On the other hand, \href{https://github.com/iden3/circom}{circom} also includes conventional variables (\verb|var|). As we use snarkjs and hence a framework for proof creation and verification based on \acp{SNARK}, a so-called trusted setup is required initially. It consists of two phases, one of which is independent of the circuit (`powers of tau'\footnote{The `powers of tau' ceremony, also referred to as `phase~1 trusted setup', is a circuit-agnostic \ac{MPC} ceremony where multiple independent parties collaboratively construct common parameters from their secret random values. The parameters allow to obtain a proving and a verification key in a later stage. Note that the \ac{SNARK} protocol's integrity guarantee (``soundness'') is compromised if all parties' random values are exposed. It is, however, important to note that while information about the private inputs to the \ac{MPC} would allow to create fake proofs and hence to violate the integrity of our artifact, the privacy of the participants' data would still be ensured even in this case~\citep{wilcox2021snark}.}) and that we could just import, as well as one that is circuit-specific. For implementation purposes, we conducted the trusted setup alone. However, for practical applications, a group of trusted parties would need to conduct an \ac{MPC} such that the participants in the \ac{FL} system would be confident that at least one group member deleted their input to the \ac{MPC}. 

We will include several, truncated source code excerpts throughout the following section. Please find the whole project on \href{https://github.com/timon131/ma_webstorm_v4}{GitHub}\footnote{\url{github.com/timon131/ma_webstorm_v4}}. Note that some templates are taken from \href{https://www.iden3.io/}{iden3}'s \href{https://github.com/iden3/circomlib}{circomlib}.

\subsubsection{Weight Proof \texorpdfstring{$\pi^w$}{}}
\label{sec:arch_implementation_weight}

We start by explaining the implementation of $\pi^w$'s circuit, namely \verb|LinRegParams(...)|. First, we define the private and public inputs according to~\eqref{eq:pi^c}. The main template execution requires various variables (cf. listing~\ref{code:weightproof_input}), some of which are also input signals to the circuit for \href{https://github.com/iden3/circom}{circom}-specific reasons. Providing an untruthful value for some of these variables might make adversarial attacks possible. To make sure that these critical input variables are correct, the circuit requires the respective values as input signals to check equality.
Listing~\ref{code:weightproof_main} provides an overview of the implementation of $\pi^w$. The circuit as the basis for the \ac{ZKP} is structured into five parts:

\begin{itemize}
    \item \textbf{Step 1} -- Range proofs for $\varepsilon_{\mu}$ and $\varepsilon_{\sigma}$:\\
    To ensure that $\mathbf{D}$'s mean $\mu \approx 0$ and variance $\sigma \approx 1$, a certain accuracy is set by $\varepsilon_{\mu}$ and $\varepsilon_{\sigma}$ respectively. Based on both values, \verb|LinRegParams| checks the accuracy of all $\mu$ and $\sigma$ for $X_1, \ldots, X_k$ and $Y$. For example, setting \verb|in_require_meanxn_acc| $= 3$ would require that the absolute value of every $\mu\:\cdot\:n$ is smaller than $\varepsilon_{\mu} = 10^{-3}$ (taking into account the conversion in~\eqref{equation:input_conversion} using $d$). Analogously, \verb|in_require_varxn| sets the upper bound for $\sigma\:\cdot\:n$ via $\varepsilon_{\sigma}$.
    \item \textbf{Step 2} -- Check $rt^{\mathbf{D}}$: \\ 
    This step rebuilds $\mathbf{D}$'s Merkle tree with one data point (a prime field element as matrix entry) at each leaf. To improve the system's performance, we do not hash the particular leaves on the lowest level since hashing is costly. This works because the hashing algorithm operates on big numbers that can be sufficiently large to cover any prime field element. After computing the tree, the template ensures that the computed root equals the public input $rt^{\mathbf{D}}$, which will be compared to the commitment specified at registration when calling the smart contract's method \verb|Clients|.
    \item \textbf{Steps 3 and 6} -- Range proofs for $\varepsilon_{\mathrm{inverse}}$ and $\varepsilon_{w^{\prime}}$: \\
    To verify the upper bound on $\varepsilon_{\mathrm{inverse}}$, the template checks the proximity of every entry in $\left(( \mathbf{X}^{\intercal} \mathbf{X}) \mathbf{Z} - \mathbbm{1} \right)$ to~$0$. For example, setting \verb|in_require_XX_acc| $= 3$ would require that the absolute value of every entry is smaller than $\varepsilon_{\mathrm{inverse}} = 10^{-3}$ (again taking into account the conversion in~\eqref{equation:input_conversion} using $d$). The same applies to \verb|in_require_b_noisy_acc|, $\varepsilon_{w^{\prime}}$, and $w^{\prime} - \tilde{w}^{\prime}$.
    \item \textbf{Steps 4 and 5} -- Range proofs for $\vartheta_{\mathbf{Z}}$ and $\vartheta_{\mathbf{X}^{\intercal} Y}$ \\
    Both $\vartheta_{\mathbf{Z}}$ and $\vartheta_{\mathbf{X}^{\intercal} Y}$ must be provided as inputs to the template in absolute numbers. The main template first finds the largest (by absolute value) entry in the matrix or vector using the maximum norm:
    \begin{align}
        \lVert \mathbf{A} \rVert = k\cdot\max\limits_{s,t} |a_{s,t}|
    \end{align}
    where $\mathbf{A}$ can be either $\mathbf{Z} \in \mathbb{N}^{\left( k+1 \right) \times \left( k+1 \right)}$ or $\mathbf{\mathbf{X}^{\intercal}Y} \in \mathbb{N}^{k+1}$. Then it checks whether \mbox{$0 \leq \lVert\mathbf{A}\rVert \leq \vartheta_{\mathbf{A}}$} holds for both choices of $\mathbf{A}$.
\end{itemize}

Note that along with these six steps, the template performs the computation of $w^{\prime}$ iteratively. The \ac{DP} noise is added in the last step by generating $h_j$ using~\eqref{equation:DP_hash_j}, choosing the respective \ac{DP} noise $q$ from $L$ using~\eqref{equation:DP_L} and~\eqref{equation:DP_p}, as well as adding $q$ as in~\eqref{equation:DP-noise}.

\subsubsection{Cost Proof \texorpdfstring{$\pi^c$}{}}
\label{sec:arch_implementation_cost}

    Next, we outline the implementation of $\pi^c$ in our proposed system, for which \verb|LinRegCost(...)| is the main template. $\pi^c$'s private inputs (cf. lines 3 to 10 in Listing~\ref{code:costproof_input}) resemble those of $\pi^w$ except for $w^+$ and $\mathrm{Sign}(w_i)$, which are required to reproduce $c$ as outlined in Section~\ref{sec:arch_conarch_cost}. Note that the inputs of $\mathbf{D}_{\mathrm{test}}$ in lines 12 to 15 of Figure~\ref{code:costproof_input} are declared as \verb|private| signals due to performance reasons (cf. Section~\ref{sec:evaluation} for details), even though they are publicly available. Also $\pi^c$ consists of five major steps to make sure that clients compute and submit their cost $c$ truthfully (cf. Listing~\ref{code:costproof_main}):
    
    \begin{itemize}
        \item \textbf{Steps 1 and 2} -- Check $rt^{\mathbf{D}}$ and $rt^{\mathbf{D}_{\mathrm{test}}}$: \\
        In addition to checking the Merkle tree root $rt^{\mathbf{D}}$ (as in \verb|LinRegProof(...)|), $\pi^c$ also ensures that clients calculate their $c$ based on the test data set $\mathbf{D}_{\mathrm{test}} = \left(\mathbf{X}_{\mathrm{test}}~Y_{\mathrm{test}}\right)$ by requiring $\tilde{rt}^{\mathbf{D}_{\mathrm{test}}} = rt^{\mathbf{D}_{\mathrm{test}}}$. $\mathbf{D}_{\mathrm{test}}$ is a standardized, publicly available, and central data set that is made available to all clients through a cloud service. Besides, $\mathbf{D}$'s standardization does not need to be checked again, as $\pi^w$ already ensures its standardization and $\pi^c$ verifies that $\tilde{rt}^{\mathbf{D}}$ equals $rt^{\mathbf{D}}$.
        \item \textbf{Step 3} -- Range proof for $\varepsilon_w$: \\
        Similar to step 5 in \verb|LinRegProof(...)|, we allow controlling $\lVert w - \tilde{w} \rVert$ by performing a range proof using the input $\varepsilon_w$. This means that essentially, the computation of the unperturbed weight $\tilde{w}$ is repeated and, subsequently, its proximity to $w$ is checked analogous to $\pi^w$'s range proof for $\varepsilon_{w^{\prime}}$.
        \item \textbf{Steps 4 and 5} -- Check $c$: \\
        First, as in \eqref{equation:cost}, step 4 estimates $\hat{Y}_{\mathrm{test}}$. Second, step 5 derives $\tilde{c}$ and ensures that $\tilde{c}$ equals the submitted $c$.
    \end{itemize}

\FloatBarrier
\subsubsection{Smart Contract}
\label{sec:arch_implementation_smartcontract}

    After introducing \verb|LinRegProof(...)| and \verb|LinRegCost(...)|, this section provides an overview of the governance structure implied by the main smart contract, namely \verb|Clients|. 
    To reduce \verb|Clients|'s size, we implemented \verb|lib| as a support library. \verb|lib| defines all structs required by \verb|Clients| (cf. Listing~\ref{code:SC_lib}) and repeatedly used methods like, e.g., proof verification calls. We will briefly introduce both major structs \verb|FL_client| and \verb|FL_generic|:
    \begin{itemize}
        \item \verb|FL_generic|: Contains all global variables for, e.g., defining $k$, $n$, $rt^{\mathbf{D}_{\mathrm{test}}}$, $L$, or the variables to control the range proofs. The client that initially deploys the \verb|Clients| smart contract must instantiate \verb|fl_generic|, which will be the only global instance of the struct.
        \item \verb|FL_client|: Upon registering, all clients are assigned to an instance \verb|fl_client| (implemented via a mapping \verb|mapclient| from the particular client's address to the respective instance \verb|fl_client|). \verb|FL_client| includes all client-specific data as, for example, $w^{\prime}_i$, $rt^{\mathbf{D}}_i$, and both proofs.
    \end{itemize}
    
    To register, clients call the smart contract's \verb|registerClient(...)| function. As depicted in Listing~\ref{code:SC_main}, calling \verb|registerClient(...)| requires $rt^{\mathbf{D}}_i$ as input, since clients must commit to their data set $\mathbf{D}_i$ upon joining the system. Further, they have to pay the admission fee $B$ defined in \verb|fl_generic| which will be used to distribute the incentive payments later. Moreover, a \verb|clientID|, a mapping \verb|mapID| to connect the clients' addresses with their \verb|clientID|, and the current block hash (at the time of registering) serving as source of public randomness for deriving their \ac{DP} noise $p$ will be set automatically. 
    
    Subsequently, clients can upload their $w^{\prime}_i$ by calling the function \verb|uploadBeta(...)| and delivering the following inputs:
    \begin{itemize}
        \item \verb|t_betaverifier|: Address of the deployed weight verifier contract. The client must deploy this contract before calling \verb|uploadBeta(...)| and provide the respective address such that other clients and \verb|uploadBeta(...)| itself can verify the client's weight proof.
        \item \verb|betaproof|: This struct (cf. Listing~\ref{code:SC_lib}) contains the actual proof, which is also the first part of the call data for verifying the weight proof.
        \item \verb|beta_noisy_true| is the struct containing the submitted $w^{\prime}_i$.
    \end{itemize}
    \verb|uploadBeta(...)| will collect the respective public inputs (cf. Listing~\ref{code:weightproof_input}) and, together with the provided \verb|betaproof|, verify the proof $\pi^w$ onchain. Successfully verifying the proof requires correct input data $\mathrm{D}_i$, the correct approximate inverse $\mathrm{Z}$ (accuracy controlled by $d$), and correct computation of $w^{\prime}_i$. If successful, \verb|uploadBeta(...)| saves the submitted \verb|beta| (i.e., the client's $w^{\prime}_i$), sets the client's \verb|betaproof_valid| to \verb|true|, and updates \verb|betaglobal| (i.e., $w_g$). The procedure is analogous (except for updating \verb|betaglobal|) when submitting the cost $c_i$ by calling \verb|uploadCost(...)|, such that we will not further describe the function.
    
    Eventually, to trigger the distribution of incentive payments, only the initial client \verb|t_initialclient| can call the function \verb|Incentivize_clients(...)|. \verb|Incentivize_clients(...)| will calculate the incentive payments $V$ and distribute them to those clients, who have successfully submitted their $w^{\prime}_i$, according to their contribution as described in Section~\ref{sec:arch_conarch_incentives}. However, as this requires trust in the initial client (as only she/he can prompt the incentive payment by calling \verb|Incentivize_clients(...)|), this way of triggering the payoff is probably not suitable in all scenarios. To make the trigger more trustless, the initial client could also specify ways to prompt the payoff. For example, any client could trigger it as soon as a certain threshold of participants is reached, or a certain amount of time (measured in the block number) has passed.
\section{Evaluation}
\label{sec:evaluation}

    To evaluate our system, we ran several tests with different parameters. $k$ and $n$ are mainly influencing the complexity of our system, since they determine the number of Merkle tree leaves and, thus, the number of hashes that we compute in a \ac{SNARK}. Note that more specifically, $(k+1)\,\cdot\,n$ for $\pi^w$ and $(k+1)\,\cdot\,(n+n_{\mathrm{test}})$ for $\pi^c$ determines the number of leaves  since we leave out $X_{0,i} = (1,\ldots,1)$ and instead only hash $X_{1,i},\ldots,X_{k,i}$ as well as $Y_i$. Besides the number of leaves, also the choice of the hashing function can have a significant impact on the system performance. For example, when switching from `Poseidon'~\citep{grassi2021poseidon} to `MiMC'~\citep{albrecht2016mimc} hashing, the number of constraints of our system roughly increases by a factor $4$. Since this factor remains approximately the same when increasing circuit complexity (i.e., the number of constraints), we tested our artifact only with `Poseidon' hashes on an AWS instance (Ubuntu 20.04, 16 virtual CPU cores, and 64 GB RAM). Instead of \href{https://github.com/iden3/snarkjs}{snarkjs}' default web assembly compiler, we used a native compiler\footnote{\href{https://github.com/Fluidex/snarkit}{snarkit by Fluidex}} (C\texttt{++}) that can handle larger circuits.
    
    Table \ref{tab:evaluation_pi-w_partial} provides an excerpt of our test results (Tables~\ref{tab:evaluation_pi-w_full}~and~\ref{tab:evaluation_pi-c} in the appendix contain the entire test data for $\pi^w$ and $\pi^c$ respectively). Note that we used \href{https://github.com/iden3/rapidsnark}{rapidsnark} to generate the proofs (cf. ``proof gen duration''). Besides, due to their similar implementation (cf. Section~\ref{sec:arch_implementation}), the cost proof's performance test results do not differ significantly from the weight proof's results. We will, therefore, mainly focus on the weight proof's results: 
    \begin{table}[!htb]
        \centering
        \scriptsize
        \setlength{\aboverulesep}{0pt}
        \setlength{\belowrulesep}{0pt}
        \renewcommand{\arraystretch}{1.5}
        \resizebox{\columnwidth}{!}{
        \begin{tabular}{|r|r|r||R{1.25cm}|R{1.25cm}|R{1.4cm}|R{1.6cm}|R{1.25cm}|R{1.6cm}|R{1.25cm}|}\toprule
        
        \textbf{$k$} & \textbf{$n$} & \textbf{$(k+1)\,\cdot\,n$} & \textbf{\# constraints} & \textbf{key gen duration (s)} & \textbf{proof gen duration (s)} & \textbf{verification duration (s)} & \textbf{circuit size (MB)} & \textbf{verification key size (kB)} & \textbf{proof size (bytes)} \\
        \midrule
        4 & 100 & 500 & 89,819 & 177 & 8.641 & 0.236 & 46 & 25 & 708 \\
        4 & 250 & 1,250 & 208,162 & 337 & 19.410 & 0.237 & 102 & 25 & 709 \\
        4 & 500 & 2,500 & 404,962 & 659 & 36.868 & 0.237 & 198 & 25 & 708 \\
        4 & 750 & 3,750 & 601,520 & 1089 & 57.042 & 0.237 & 310 & 25 & 705 \\
        4 & 1,000 & 5,000 & 798,659 & 1360 & 74.945 & 0.232 & 391 & 25 & 705 \\
        4 & 1,500 & 7,500 & 1,190,620 & 2571 & 120.736 & 0.263 & 614 & 25 & 705 \\
        \midrule
        \end{tabular}
        }
        \caption{Evaluation of essential parameters for building and verifying the $\pi^w$ circuit depending on $k$ and $n$.}
        \label{tab:evaluation_pi-w_partial}
    \end{table}
    Before analyzing our system's performance, we describe an essential optimization that prior tests suggested: We found that the recomputations of the Merkle trees are the artifact's dominating operations in terms of constraints. Therefore, we optimized the computation of $rt^{\mathbf{D}}$ and $rt^{\mathbf{D}_{\mathrm{test}}}$ by hashing six instead of two data points at the leaf level at once and hard-coding `empty' leaf hashes to avoid additional hashing in those Merkle trees that are not entirely filled with data points on the leaf level (depending on $k$, $n$, and $n_{\mathrm{test}}$, not every leaf necessarily represents a data point as the bottom level must -- given our Merkle tree optimization -- always contain $2^{\,\mathrm{depth}}\,\cdot\,3$ leaves). This optimization reduced the circuit complexity by roughly 50\%. Then, computing the circuit-specific trusted setup for a \ac{LR} on a training data set with $k=4$ features and sample size $n=1,000$ per client in a \ac{FL} setting takes roughly $23$ minutes each for $\pi^w$ (cf. Table~\ref{tab:evaluation_pi-w_partial}) and $\pi^c$ (cf. Table~\ref{tab:evaluation_pi-c} in the appendix). Note that, as outlined above, the circuit-specific trusted setup must only be computed once and can then be used by all clients for a particular \ac{FL} learning task. 
    Using the proving key from the trusted setup, every client must spend roughly $2.5$ minutes on computing both proofs that allow fellow clients to verify their submitted weight $w$ and cost $c$ (for $k=4$ and $n=1,000$). In practical \ac{FL} applications, the proof generation likely runs on less sophisticated machines than our AWS instance. We, therefore, also tested the proof duration on a single CPU core (also for combinations other than $k=4$ and $n=1,000$). The results reveal that the witness generation duration was not sensitive to this limitation, whereas the duration for generating the proof using \href{https://github.com/iden3/rapidsnark}{rapidsnark} increased seven times from $4.3$ to $29.4$ seconds and the proof verification duration four times from $0.24$ to $0.97$ seconds. In total, the proof duration grew $1.4$ times from around $75$ to $103$ seconds. Moreover, the RAM required to generate and validate proofs remains below 4 GB and agnostic to the number of constraints.
    Looking at the performance of the system's smart contract on the \ac{EVM}, we find gas cost of around $2.2$~million for all $\pi^w$-related on-chain operations and $920,000$ for all $\pi^c$-related operations. This translates into a total of roughly USD~$1270$ for all on-chain operations (USD~$900$ for $\pi^w$ operations and USD~$370$ for $\pi^c$ operations) per client, given a gas price of 100~gwei and a rate of 1~ETH~=~USD~4,000. While in absolute terms this is prohibitively expensive, it is only $21\%$ of the current public Ethereum target block capacity as well as $10.5\%$ of the block limit. Note that the $\pi^c$ operations are significantly cheaper than the $\pi^w$ operations, mainly since the latter requires the vector $L$ as public input and updates $w_{\mathrm{g}}$. As long as operations on public blockchains are that costly, using a permissioned blockchain like Quorum can allow not only to reduce costs but also allowing for considerably higher throughput~\citep{sedlmeir2021benchmarking}. 
    
\begin{figure}[!htb]
        \begin{subfigure}{.5\textwidth}
            \centering
            \includegraphics[page=1, width=\linewidth, trim=10.9cm 5.5cm 10.9cm 5.5cm, clip]{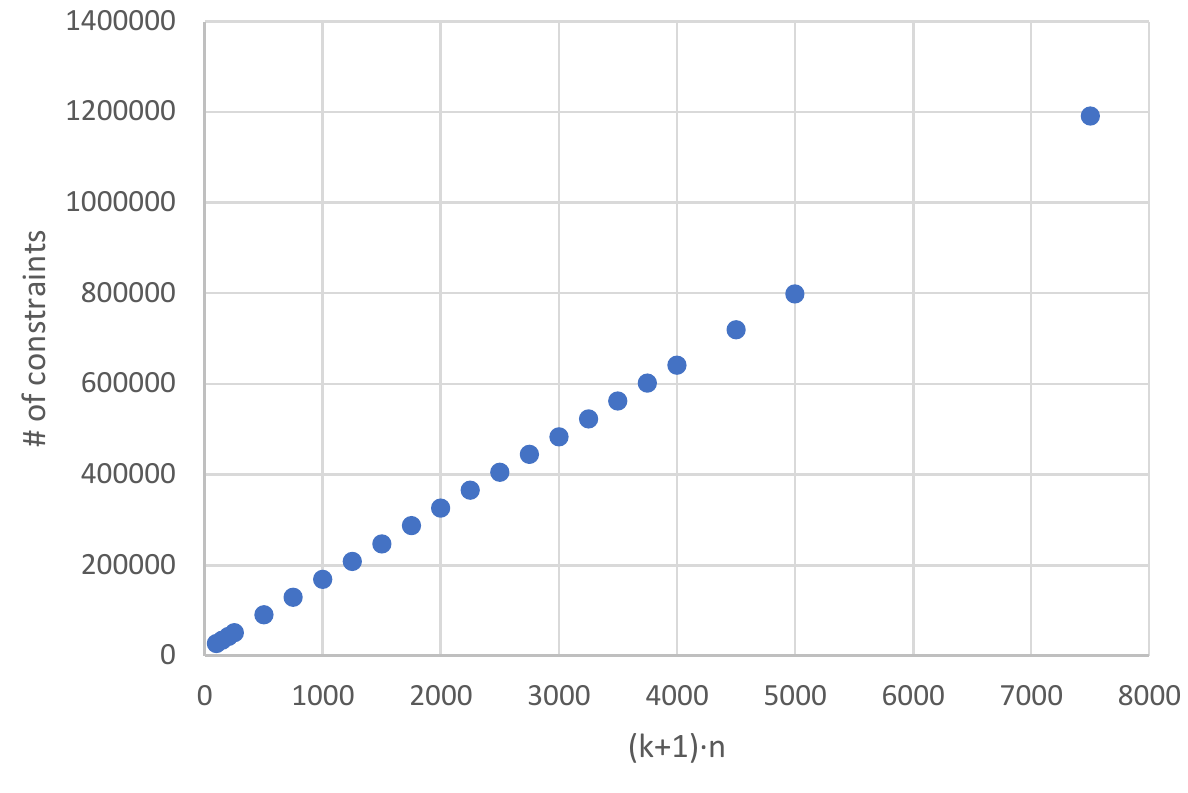}
            \caption{Scaling of the circuit complexity (i.e., the number of constraints) with $(k+1)\,\cdot\,n$.}
            \label{plot:scalability_pi-w_constraints}
        \end{subfigure}
        \begin{subfigure}{.5\textwidth}
            \centering
            \includegraphics[page=5, width=\linewidth, trim=10.9cm 5.5cm 10.9cm 5.5cm, clip]{Figures/Performance-plots_v3.pdf}
            \caption{Scaling of the verification duration, verification key size, and proof size with the number of constraints.}
            \label{plot:scalability_pi-w_constant}
        \end{subfigure}
        \caption{$\pi^w$ scalability analysis (client perspective).}
        \label{plot:scalability_pi-w}
    \end{figure}

    To assess our system's scalability, we focus on analyzing the impacts of increasing $k$, $n$, and $n_{\mathrm{test}}$ on the circuit-specific trusted setup, the proof generation and verification, as well as the gas cost of the smart contract. Both Figure~\ref{plot:scalability_pi-w} and Table~\ref{tab:evaluation_pi-w_partial} show results that are in line with the expected scaling properties of \acp{SNARK}: The computation of the circuit-specific trusted setups (i.e., `key gen duration') as well as the time it takes to generate the proofs (i.e., `proof duration') scale linearly with \mbox{$(k+1)\,\cdot\,n$} or \mbox{$(k+1)\,\cdot\,(n+n_{\mathrm{test}})$} (cf. Table~\ref{tab:evaluation_pi-c} in the appendix). Moreover, verification duration, verification key size, and proof size do not significantly change when increasing the circuit complexity. Thus, increasing $k$, $n$, or $n_{\mathrm{test}}$ will have only minor effects on the system's overall computation effort, since the proof duration might grow for an individual client whilst the redundant on-chain proof storage and verification cost will remain constant. Even when translating our results to circuits as huge as those in the \href{https://github.com/hermeznetwork}{Hermez project} (the project also used \href{https://github.com/iden3/circom}{circom} with \href{https://github.com/iden3/snarkjs}{snarkjs} and manages to compute a proof for around $100$~million constraints in few minutes on a server with 64 virtual cores and 1 TB RAM~\citep{hermez2021rapidsnark}), we find promising scaling performance: Given the test results, we expect that $\pi^w$ and $\pi^c$ with \mbox{$(k+1)\,\cdot\,n = 600,000$} data points and \mbox{$n_{\mathrm{test}} = 0.1\,\cdot\,n = 6,000$} would result in roughly $100$ million constraints and hence be feasible to prove in a few minutes to hours per proof (depending on the hardware used; in our setup roughly $2.5$ hours with 16~CPU cores or approximately~$3.5$ hours with one core) whilst handling a huge sample size (e.g.,~$k=5$ and $n=100,000$). In this case, the one-time trusted setups would take a few hours to a few days for each participant of the trusted setup (which we assume will not necessarily be conducted by the clients in practice, but a small to medium-sized group or maybe also research institutions, as could be observed for the trusted setup used in Z-Cash, zkSync, etc.). Besides, we expect the verification duration, verification key size, and proof size to remain constant in this scenario.
    
    Also, the gas cost for the \ac{EVM} operations is independent of the proof complexity. The only data that must be stored on-chain to verify a proof are the proof itself, the verification key, and the public inputs. Both the proof size as well as the verification key size are constant for every $k$, $n$, and $n_{\mathrm{test}}$. The public inputs' volume only increases when either the number of features $k$ or the \ac{DP} noise discretization interval $d_{\mathcal{L}}$ grows, since this leads to an increase in the number of entries in the weight vector $w^{\prime}$ or the \ac{DP} noise vector $L$. Thus, raising $w^{\prime}$ or $L$ increases the payload size, which in turn leads to a rise in gas cost. However, for moderate $k$ and $d_{\mathcal{L}}$, the total gas cost remains small since the respective proof verification itself is responsible for the major part of the gas cost. Therefore, both proofs can be stored as well as verified cheaply on-chain also for large circuits and the particular gas cost remains, as mentioned above, approximately at $2.2$~million for all $\pi^w$-related on-chain operations and $920,000$ for all $\pi^c$-related operations.
    Eventually, a clients' effort does not grow with the size of other clients' underlying data sets or the number of participating clients. Further, the number of on-chain transactions only grows linearly with the number of clients and is independent of the size of the clients' training data as well as the central test data. These results promise high potential also for more sophisticated \ac{ML} applications that require higher data volumes.
    
    Eventually, we refrain from testing the accuracy of our \ac{FL} system. The only difference regarding accuracy between our and a plain \ac{FL} system is the \ac{DP} noise that we add to the clients’ weights; and since aggregation is an average over all local weights plus noise, this difference is just an average of \ac{IID} Laplacian-distributed random variables, which has mean $0$ and standard deviation proportional to $\frac{\varepsilon}{\sqrt{\lvert I\rvert}}$. The value of $\varepsilon$ and $\lvert I\rvert$ hence completely determine the change in accuracy of our approach compared to the literature.
\section{Discussion and Conclusion}
\label{sec:conclusion}
    
Our paper aims at answering the research question of how a \ac{FL} system can achieve fairness, integrity, and privacy simultaneously whilst still being practical and scalable. After identifying a business need for these requirements and arguing that related work so far has not proposed an architecture that satisfies them, we described our proposed system in Section~\ref{sec:architecture} based on a combination of blockchain, \ac{ZKP}, and \ac{DP}. Our conceptual discussion of the architecture and the experiments that we describe in Section~\ref{sec:evaluation} suggest that our implementation indeed offers a practical solution to confidential, fair, and tamper-resistant \ac{FL} that achieves reasonable performance and scalability properties for \acp{LR}. Next to suggesting a pathway to effectively combining \ac{FL} with blockchain, \ac{ZKP}, and \ac{DP}, we develop a system that allows clients to verifying that other clients truthfully trained their local model and received a fair compensation for participating in the case of
\ac{LR}. Thus, we go beyond existing research that has suggested \acp{ZKP} for verifying \ac{ML} model inference based on already trained models.

However, our research is not without limitations and reveals potential for future research that we will outline in this section to conclude our paper. First, even though our suggested system does currently only support \acp{LR}, we tried to design it as generic as possible to allow for adapting the system to other classes of \ac{ML} protocols. Specifically, we move computationally intensive parts (in our case, the inversion of $\mathbf{X}^\intercal\mathbf{X}$) outside the circuit and only check certain properties of the result (in our case, that the approximate inverse that must be provided as private input to the \ac{ZKP} is indeed close to the true inverse, cf. Section~\ref{sec:arch_conarch_parameters}) rather than recomputing the result. For the case of \ac{LR}, we derived an error bound (see~\eqref{eq:error_bound_inv}) that allowed us to significantly reduce the complexity of the respective circuit while maintaining full tamper resistance. The approach of only proving specific properties that the local weights need to satisfy to indicate the integrity of training may be generalizable to further classes of \ac{ML} models, also considering the recent advancement in using \acp{ZKP} for frequent operations in \ac{ML} (e.g., \citep{weng2021mystique}). Since training \ac{ML} models often involves solving a convex optimization problem where optimality can be checked locally (e.g., all partial derivatives in the allowed directions are $0$)~\citep{bubeck2014convex}, we encourage future research to adapt the suggested architecture to more sophisticated classes of \ac{ML} models beyond \ac{LR}. Furthermore, the aggregation of weights in federated learning is often linear~\citep{Nilsson_2018_Performance}, so our concept of perturbing the local weights with verifiable noise is applicable beyond multiple \acp{LR}: Assuming that the training data is \ac{IID} (a common assumption in \ac{FL} applications), the local weights are also \ac{IID}. Moreover, as we construct the clients' \ac{DP} noise from a random oracle, the noise is also \ac{IID} and has mean $0$ and finite variance by construction. Since the weights and the noise applied to them are by construction independent, by the linearity of the averaging algorithm FedAvg, the error in the global model is a weighted average of local noise and the random variable (noise times weight) is \ac{IID} with expectation value $0$ and finite variance. By the law of large numbers, the error in the aggregated global model hence converges to $0$. Thus, for a large number of clients, the error term in the aggregate model as introduced by \ac{DP} is typically small. It follows that our approach of using \ac{DP} (which has been proposed by several other scholars but is particularly relevant in our system because we can prove not only the correct training but also the correct addition of noise) extends to more sophisticated \ac{ML} models beyond \ac{LR}.

To further improve the performance and practicality of our system, we aim at implementing the \acp{ZKP} via \acp{STARK} for improved proof creation performance, post-quantum security, and eliminating the need for a trusted setup in the future. Further, the system's scalability would further benefit from a recursive verification mechanism that reduces the complexity of verifying weight and cost proofs. This could be implemented by building on batching techniques~\citep{gailly2021snarkpack} or recursive proofs~\citep{bowe2020zexe,chiesa2020fractal} and would facilitate scalability by hierarchical aggregation (e.g., only $\log(\lvert I\rvert)$ verification steps would be required on-chain). Besides, even though recalculating $w$ in $\pi^c$ is not the proof's main complexity driver, we see optimization potential there: One could commit a hash of the weight (without noise) and some random salt in the smart contract (proving that the hash was computed like that), so for computing the costs we only need to prove that we used the pre-image of this hash (without salt) for the computation. This can further reduce the cost proof's number of constraints. Next to optimizations of our own code and the performance improvements in libraries that support the generation of \ac{ZKP} (e.g., \acp{STARK} or \href{https://github.com/iden3/circom}{circom} 2.0), we are also confident that hardware acceleration for faster \ac{ZKP}-related operations and particularly proving will be available soon, as research in this area is already conducted, e.g., by projects that build on Ethereum and that leverage \acp{ZKP}~\citep{matterlabs2020acceleration}. This may allow getting even shorter proof times also on devices that are computationally more restricted than a Laptop. 

Moreover, we acknowledge that there are still some attacks on integrity: When clients know the learning task (i.e., in our case, the parameters stored on \verb|fl_generic|) prior to committing their data when joining the system, they could attack the system by manipulating their data set before committing to it. However, reverse engineering the data in order to get a desired (malicious) result for the weights is likely more effort than just contributing arbitrarily chosen weights. Moreover, we expect many use cases for our proposed system to be built on sensor data (e.g., from vehicular networks or health applications). Given this, and the availability of certified sensors (e.g., by means of a crypto-chip on the sensor and a certificate of the manufacturer), as are emerging, for example, in Germany's Smart Meter rollout~\citep{djamali2021asset}, the \ac{ZKP}-based approach could handle this issue by including a proof of authenticity (i.e., a proof that the data was signed by a private key that is bound to a certificate that was in turn signed by a trusted, publicly known entity) for the sensor input data when committing to it. This would be easy to integrate at the costs of an additional signature verification per Merkle tree leaf (around 5,000 constraints per leaf for a Schnorr signature). This still does not protect against physical manipulation (imagine putting a temperature sensor to a place where it is not supposed to be), but may offer a reasonable degree of trust in data provenance in many practical scenarios.

Our current incentive mechanism relies on the existence of a central and public data set as described in Section~\ref{sec:arch_conarch_incentives}. Since we acknowledge that this hypothesis is not always feasible in practice, we will work on developing other, effective incentive mechanisms. In doing so, we intend to ensure the mechanisms' fairness by incorporating the research of~\citet{Shapley1953Value}. In doing so, our construction allows us to perform the local evaluation of each client on their actual weights with and without noise, so we get new degrees of freedom for proving the correct evaluation and deriving fair incentives.

Lastly, to ensure the system's applicability in practice, we plan a twofold approach. First, as we are currently training the test model on the \href{https://www.kaggle.com/camnugent/california-housing-prices}{California Housing Prices} data set, we will also run accuracy tests to assess how our conceptual approach translates into data-related performance in practice. Second, we will evaluate the perspective of enterprises on the new features that the combination of \ac{FL} with blockchain technology, \ac{DP}, and \ac{ZKP} offers.

We believe that the convergence of emerging technologies like \ac{ML} and blockchains in the context of data generated by the IoT, as \citet{guggenberger2021emerging} or \citet{singh2020blockiotintelligence} suggest, combined with privacy-enhancing technologies in the context of sensor data sharing or derived models in data markets~\citep{munilla2021revealing}, has the potential to facilitate many new use cases as well as business models and can inspire the field of computer science. As all these aspects are relevant for \ac{FL}, we are expecting interesting results that further extend our design to more complex models in the future.
~\\~\\\textbf{Acknowledgment:}
We want to thank Orestis Papageorgiou, who pointed us to the work of~\citet{Newman1974How}, and Jordi Baylina and the \href{https://iden3.io/}{iden3}-Team for their great work on \href{https://github.com/iden3/circom}{circom} and \href{https://github.com/iden3/snarkjs}{snarkjs} and their helpful responses to our questions. Moreover, we thank Matthias Babel, Jonas Berweiler, Dennis Jelito, and  Michael Reichle for their ad-hoc problem-solving support. We also gratefully acknowledge the Bavarian Ministry of Economic Affairs, Regional Development and Energy for their funding of the project ``Fraunhofer Blockchain Center (20-3066-2-6-14)'' that made this paper possible.


\bibliographystyle{unsrtnat}
\bibliography{99_Literature.bib}

\clearpage
\section*{Appendix}

\subsection{Selected Code Snippets}
    \label{sec:appendix}
    \begin{lstlisting}[language=circom, basicstyle=\scriptsize\ttfamily, escapeinside={(*}{*)}, caption={Excerpt from the circuit for the weight proof $\pi^w$ -- Input signals. The default visibility for inputs is public.}, label={code:weightproof_input}]
template LinRegParams(k, n, dec, l_train, require_meanxn_acc, require_varxn_acc, require_XX_acc, require_XX_inv_maxnorm, require_X_trans_Y_maxnorm, require_b_noisy_acc, hash_alg, DP_acc) {
    
    signal private input in_x_pos[k][n];            (*\textcolor{gray}{// $(\mathbf{X}_i)^+$}*)
    signal private input in_x_sign[k][n];           (*\textcolor{gray}{// $\mathbf{Sign(X)}_i$}*)
    signal private input in_y_pos[n][1];            (*\textcolor{gray}{// $(Y_i)^+$}*)
    signal private input in_y_sign[n][1];           (*\textcolor{gray}{// $\mathrm{Sign}(Y_i)$}*)
    signal private input in_xx_inv_pos[k][k];       (*\textcolor{gray}{// $(\mathbf{Z}_i)^+$}*)
    signal private input in_xx_inv_sign[k][k];      (*\textcolor{gray}{// $\mathbf{Sign(Z)}_i$}*)
    //public inputs:
    signal input in_k;                              (*\textcolor{gray}{// $k$}*)
    signal input in_n;                              (*\textcolor{gray}{// $n$}*)
    signal input in_dec;                            (*\textcolor{gray}{// $d$}*)
    signal input in_merkleroot_train;               (*\textcolor{gray}{// $rt^{\mathbf{D}}_i$}*)
    signal input in_Lap_X_pos[DP_acc - 1];          (*\textcolor{gray}{// $L$}*)
    signal input in_DP_acc;                         (*\textcolor{gray}{// $d_{\mathcal{L}}$}*)
    signal input in_hash_BC;                        (*\textcolor{gray}{// $\mathrm{block~hash~or~other~source~of~public~randomness}$}*)
    signal input in_b_noisy_true_pos[k][1];         (*\textcolor{gray}{// $(w^{\prime}_i)^+$}*)
    signal input in_b_noisy_true_sign[k][1];        (*\textcolor{gray}{// $\mathrm{Sign}(w^{\prime})_i$}*)
    signal input in_require_meanxn_acc;             (*\textcolor{gray}{// $\sim \varepsilon_{\mu}$}*)
    signal input in_require_varxn_acc;              (*\textcolor{gray}{// $\sim \varepsilon_{\sigma}$}*)
    signal input in_require_XX_acc;                 (*\textcolor{gray}{// $\sim \varepsilon_{\mathrm{inverse}}$}*)
    signal input in_require_XX_inv_maxnorm;         (*\textcolor{gray}{// $\vartheta_{\mathbf{Z}}$}*)
    signal input in_require_X_trans_Y_maxnorm;      (*\textcolor{gray}{// $\vartheta_{\mathbf{X}^{\intercal} Y}$}*)
    signal input in_require_b_noisy_acc;            (*\textcolor{gray}{// $\sim \varepsilon_{w^{\prime}}$}*)
    
    ...
\end{lstlisting}
    \begin{lstlisting}[language=circom, basicstyle=\scriptsize\ttfamily, escapeinside={(*}{*)}, caption={Excerpt from the circuit for the weight proof $\pi^w$ -- Main part.}, label={code:weightproof_main}]
    ...
    
    // 1. step | Check data normalization (i.e., (*\textcolor{gray}{$\varepsilon_{\mu} \approx 0$ and $\varepsilon_{\sigma} \approx 1$)}*)
    component check_mean = Check_MeanXY(k, n, dec, require_meanxn_acc);
    ...
    //make sure that range_meanxn_acc.out is within the bound (*\textcolor{gray}{$\varepsilon_{\mu}$}*)
    component range_meanxn_acc = GreaterEqThan(...);
    ...
    1 === range_meanxn_acc.out;
    ...
    component check_var = Check_VarXY(k, n, dec, ..., require_varxn_acc);
    ...
    //make sure that range_varxn_acc.out is within the bound (*\textcolor{gray}{$\varepsilon_{\sigma}$}*) 
    component range_varxn_acc = GreaterEqThan(bits_range_varxn_acc);
    ...
    1 === range_varxn_acc.out;
    
    
    // 2. step | Check (*\textcolor{gray}{$rt^{\mathbf{D}}_i$}*)
    component merkleproof_train = MerkleProof_six(k, n, l_train, hash_alg);
    ...
    //make sure that (*\textcolor{gray}{$rt^{\mathbf{D}}_i$}*) is correct
    in_merkleroot_train === merkleproof_train.out;
    ...
    
    // 3. step | range proof (*\textcolor{gray}{$\varepsilon_{\mathrm{inverse}}$}*)
    component XX_rangeproof = XX_RangeProof(k, n, require_XX_acc, dec);
    ...
    //make sure that range_XX_acc.out is within the bound (*\textcolor{gray}{$\varepsilon_{\mathrm{inverse}}$}*)
    component range_XX_acc = GreaterEqThan(...);
    ...
    1 === range_XX_acc.out;
    
    // 4. step | range proof (*\textcolor{gray}{$\vartheta_{\mathbf{Z}}$}*)
    component maxelement_XX_inv_pos = NormMaxElement(k, k, ...);
    ...
    //make sure that (*\textcolor{gray}{$k\:\cdot\:range\_XX\_inv\_norm.out$ is within the bound $\vartheta_{\mathbf{Z}}$}*)
    component range_XX_inv_norm = LessEqThan(...);
    1 === range_XX_inv_norm.out;
    
    // 5. step | range proof (*\textcolor{gray}{$\vartheta_{\mathbf{X}^{\intercal} Y}$}*)
    //compute (*\textcolor{gray}{$\mathbf{X}_i^{\intercal} Y_i$}*)
    component X_trans_Y_mult = MatrixMult(n, k, 1);
    ...
    component maxelement_X_trans_Y_pos = VectorNormMaxElement(k, ...);
    ...
    //make sure that (\textcolor{gray}{$*k\:\cdot\:$maxelement\_X\_trans\_Y\_pos.out$ is within the bound $\vartheta_{\mathbf{X}^{\intercal} Y}$}*)
    component range_X_trans_Y_norm = LessEqThan(...);
    1 === range_X_trans_Y_norm.out;
    
    // 6. step | range proof (*\textcolor{gray}{$\varepsilon_{w^{\prime}}$}*)
    component b_rangeproof_noisy = b_noisy_RangeProof(k, n, require_b_noisy_acc, hash_alg, 
    dec, DP_acc);
    ...
    //make sure that range_b_noisy_acc.out is within the bound (*\textcolor{gray}{$\varepsilon_{w^{\prime}}$}*)
    component range_b_noisy_acc = GreaterEqThan(require_b_noisy_acc);
    1 === range_b_noisy_acc.out;
}
component main = LinRegParams(...);
\end{lstlisting}
    
    \begin{lstlisting}[language=circom, basicstyle=\scriptsize\ttfamily, escapeinside={(*}{*)}, caption={Excerpt from the circuit for the cost proof $\pi^c$ -- Input signals. The default visibility for inputs is public.}, label={code:costproof_input}] 
template LinRegCost(k, n, n_test, dec, l_train, l_test, hash_alg, require_b_acc) {

    signal private input in_x_pos[k][n];                (*\textcolor{gray}{// $(\mathbf{X}_i)^+$}*)
    signal private input in_x_sign[k][n];               (*\textcolor{gray}{// $\mathbf{Sign(X)}_i$}*)
    signal private input in_y_pos[n][1];                (*\textcolor{gray}{// $(Y_i)^+$}*)
    signal private input in_y_sign[n][1];               (*\textcolor{gray}{// $\mathrm{Sign}(Y_i)$}*)
    signal private input in_b_true_pos[k][1];           (*\textcolor{gray}{// $(w_i)^+$}*)
    signal private input in_b_true_sign[k][1];          (*\textcolor{gray}{// $\mathrm{Sign}(w_i)$}*)
    signal private input in_xx_inv_pos[k][k];           (*\textcolor{gray}{// $(\mathbf{Z}_i)^+$}*)
    signal private input in_xx_inv_sign[k][k];          (*\textcolor{gray}{// $\mathbf{Sign(Z)}_i$}*)
    //(*\textcolor{gray}{$\mathbf{D}^{\mathrm{test}}$}*) is not set as public to reduce onchain data volume
    signal private input in_x_test_pos[k][n_test];      (*\textcolor{gray}{// $(\mathbf{X}_{\mathrm{test}})^+$}*)
    signal private input in_x_test_sign[k][n_test];     (*\textcolor{gray}{// $\mathbf{Sign(X)}_{\mathrm{test}}$}*)
    signal private input in_y_test_pos[n_test][1];      (*\textcolor{gray}{// $(Y_{\mathrm{test}})^+$}*)
    signal private input in_y_test_sign[n_test][1];     (*\textcolor{gray}{// $\mathrm{Sign}(Y)_{\mathrm{test}}$}*)
    //public inputs:
    signal input in_cost_submitted;                     (*\textcolor{gray}{// $c_i$}*)
    signal input in_k;                                  (*\textcolor{gray}{// $k$}*)
    signal input in_n;                                  (*\textcolor{gray}{// $n$}*)
    signal input in_n_test;                             (*\textcolor{gray}{// $n_{\mathrm{test}}$}*)
    signal input in_dec;                                (*\textcolor{gray}{// $d$}*)
    signal input in_merkleroot_train;                   (*\textcolor{gray}{// $rt^{\mathbf{D}}_i$}*)
    signal input in_merkleroot_test;                    (*\textcolor{gray}{// $rt^{\mathbf{D}_{\mathrm{test}}}$}*)
    signal input in_require_b_acc;                      (*\textcolor{gray}{// $\sim \varepsilon_{w}$}*)
    
    ...
\end{lstlisting}
    \begin{lstlisting}[language=circom, basicstyle=\scriptsize\ttfamily, escapeinside={(*}{*)}, caption={Excerpt from the circuit for the cost proof $\pi^c$ -- Main part.}, label={code:costproof_main}]
    ...
    
    // 1. step | Check (*\textcolor{gray}{$rt^{\mathbf{D}}_i$}*)
    component merkleproof_train = MerkleProof_six(k, n, l_train, hash_alg);
    ...
    //make sure that (*\textcolor{gray}{$rt^{\mathbf{D}}_i$}*) is correct
    in_merkleroot_train === merkleproof_train.out;

    // 2. step | Check (*\textcolor{gray}{$rt^{\mathbf{D}_{\mathrm{test}}}$}*)
    component merkleproof_test = MerkleProof_six(k, n_test, l_test, hash_alg);
    ...
    //make sure that (*\textcolor{gray}{$rt^{\mathbf{D}_{\mathrm{test}}}$}*) is correct
    in_merkleroot_test === merkleproof_test.out;

    // 3. step | range proof (*\textcolor{gray}{$w_i$}*)
    component b_rangeproof = b_RangeProof(k, n, require_b_acc, hash_alg, dec);
    ...
    //make sure that range_b_true_acc.out is within the bound (*\textcolor{gray}{$\varepsilon_{w}$}*)
    component range_b_true_acc = GreaterEqThan(...);
    1 === range_b_true_acc.out;

    // 4. step | compute (*\textcolor{gray}{$\hat{Y}_{\mathrm{test},i}$}*)
    //compute (*\textcolor{gray}{$\mathbf{X}^{\intercal}$}*)
    signal x_test_trans_pos[n_test][k];
    signal x_test_trans_sign[n_test][k];
    ...
    //calculate (*\textcolor{gray}{$\hat{Y}_{\mathrm{test},i}$}*)
    component y_est_mult = MatrixMult(k, n_test, 1);
    ...

    // 5. step | compute (*\textcolor{gray}{$\tilde{c}_i$}*)
    ...
    signal y_est[n_test];
    signal y_test[n_test];
    signal y_test_tmp[n_test];
    component cost_sum = SigSum(n_test);
    ...
    //make sure that (*\textcolor{gray}{$c_i = \tilde{c}_i$}*)
    in_cost_submitted === cost_sum.out;
}
\end{lstlisting}
    
    \begin{lstlisting}[language=Solidity, basicstyle=\scriptsize\ttfamily, escapeinside={(*}{*)}, caption={Excerpt from the client smart contract -- Support library.}, label={code:SC_lib}]
library lib {
    ...
    // beta
    struct Beta {
        uint[] beta_pos;                (*\textcolor{gray}{// $(w^{\prime}_i)^+$}*)
        uint8[] beta_sign;              (*\textcolor{gray}{// $\mathrm{Sign}(w^{\prime})_i$}*)
    }
    struct BetaGlobal {
        Beta beta;                      (*\textcolor{gray}{// $w_{\mathrm{g}}$}*)
        uint16 I_round;                 // number of participants with valid (*\textcolor{gray}{$\pi^w_i$}*)
    }
    struct Proof {
        uint[2] a;
        uint[2][2] b;                   // proof.json
        uint[2] c;
    }
    
    // cost struct
    struct FL_cost {
        uint[] cost;                    (*\textcolor{gray}{// $C$}*)
        address payable[] t_cost;
    }
    
    // client struct
    struct FL_client {
        uint16 clientID;
        uint merkleroot_train;          (*\textcolor{gray}{// $rt^{\mathbf{D}_i}$}*)
        uint hash_BC;                   (*\textcolor{gray}{// current block hash}*)
        
        Beta beta_noisy_true;           (*\textcolor{gray}{// $(w^{\prime}_i)$}*)
        address t_betaverifier;
        Proof betaproof;
        bool betaproof_valid;
        
        uint cost;                      (*\textcolor{gray}{// $c_i$}*)
        address t_costverifier;
        Proof costproof;
    }
        
    // generic struct
    struct FL_generic {
        uint admission_fee;             (*\textcolor{gray}{// $k$ in wei}*)
        uint8 k;                        (*\textcolor{gray}{// $k$}*)
        uint n;                         (*\textcolor{gray}{// $n$}*)
        uint n_test;                    (*\textcolor{gray}{// $n_{\mathrm{test}}$}*)
        uint8 dec;                      (*\textcolor{gray}{// $d$}*)
        uint DP_acc;                    (*\textcolor{gray}{// $d_{\mathcal{L}}$}*)
        uint merkleroot_test;           (*\textcolor{gray}{// $rt^{\mathbf{D}_{\mathrm{test}}}$}*)
        uint[] Lap_X_pos;               (*\textcolor{gray}{// $(L)^+$}*)
        uint8 require_meanxn_acc;       (*\textcolor{gray}{// $\sim \varepsilon_{\mu}$}*)
        uint8 require_varxn_acc;        (*\textcolor{gray}{// $\sim \varepsilon_{\sigma}$}*)
        uint8 require_XX_acc;           (*\textcolor{gray}{// $\sim \varepsilon_{\mathrm{inverse}}$}*)
        uint require_XX_inv_maxnorm;    (*\textcolor{gray}{// $\vartheta_{\mathbf{Z}}$}*)
        uint require_X_trans_Y_maxnorm; (*\textcolor{gray}{// $\vartheta_{\mathbf{X}^{\intercal} Y}$}*)
        uint8 require_b_noisy_acc;      (*\textcolor{gray}{// $\sim \varepsilon_{w^{\prime}}$}*)
        uint8 require_b_acc;            (*\textcolor{gray}{// $\sim \varepsilon_{w}$}*)
        uint8 hash_alg;                 (*\textcolor{gray}{// $hash\_alg$}*)
    }
    ...
}
\end{lstlisting}
    \begin{lstlisting}[language=Solidity, basicstyle=\scriptsize\ttfamily, escapeinside={(*}{*)}, caption={Excerpt from the client smart contract -- Main part.}, label={code:SC_main}]
...
contract Clients {
    
    //generic variables
    uint16 public count;
    uint32 constant betaproof_publicinput_length = 119;
    address public t_initialclient;
    uint[] public incentives;
    
    
    // initialize structs
    lib.FL_generic fl_generic;
    lib.BetaGlobal betaglobal;
    lib.FL_cost fl_cost;
    
    // mappings
    mapping (address => lib.FL_client) public mapclient;
    mapping (uint => address) public mapID;
    
    constructor(lib.FL_generic memory _fl_generic) {
        count = 0;
        t_initialclient = msg.sender;
        fl_generic = _fl_generic;
    }
    
    
    //
    // functions
    //
    
    function registerClient(uint _merkleroot_train) public payable {
        count++;
        mapID[count] = msg.sender;
        mapclient[msg.sender].clientID = count;
        mapclient[msg.sender].merkleroot_train = _merkleroot_train;
        mapclient[msg.sender].hash_BC = uint(blockhash(block.number));
        
        // require that fee has been paid
        require(msg.value >= fl_generic.admission_fee, "Pay fee");
    }
    
    function uploadBeta(lib.Proof memory _betaproof, lib.Beta memory _beta_noisy_true, address _t_betaverifier) public {
        // upload proof
        mapclient[msg.sender].t_betaverifier = _t_betaverifier;     // address of deployed verifier SC 
        mapclient[msg.sender].betaproof = _betaproof;
        
        // upload beta
        mapclient[msg.sender].beta_noisy_true = _beta_noisy_true;
        
        // verify beta proof
        bool proof = lib.verifyBeta(fl_generic, mapclient[msg.sender]);
        require (proof, "proof failed");            // no beta will be stored if proof fails
        mapclient[msg.sender].betaproof_valid = proof;
        
        // update betaglobal
        lib.Beta[] memory beta_all_valid = new lib.Beta[](count);
        uint16 i_valid = 0;
        for (uint16 i = 1; i <= count; i++) {
            if (mapclient[getClientAddress(i)].betaproof_valid) {
                beta_all_valid[i_valid] = mapclient[getClientAddress(i)].beta_noisy_true;
                i_valid++;
            }
        }
        betaglobal = lib.genBetaMean(beta_all_valid);
    }
    
    function uploadCost(lib.Proof memory _costproof, uint _cost, address _t_costverifier) public {
        // make sure that client contributed 
        require(mapclient[msg.sender].betaproof_valid, "no beta");
        
        // upload proof
        mapclient[msg.sender].t_costverifier = _t_costverifier;     // address of deployed verifier SC 
        mapclient[msg.sender].costproof = _costproof;
        
        // upload cost
        mapclient[msg.sender].cost = _cost;
        
        // verify cost proof
        bool proof = lib.verifyCost(fl_generic, mapclient[msg.sender]);
        require(proof, "proof failed");             // no cost will be stored if proof fails
        
        // upload to fl_cost
        fl_cost.cost.push(_cost);
        fl_cost.t_cost.push(payable(msg.sender));
    }
    
    function Incentivize_clients () public payable {
        // make sure that only initial client can trigger incentive distribution
        require(msg.sender == t_initialclient, "only initclient");
        
        // get incentives
        incentives = lib.calcIncentives(fl_cost.cost, fl_generic.admission_fee);
        
        // pay incentives
        require(address(this).balance >= incentives.length * fl_generic.admission_fee, "low balance");
        for (uint16 i = 0; i < count; i++) {
            fl_cost.t_cost[i].transfer(incentives[i]);
        }
    }
    ...
}
\end{lstlisting}

\pagebreak
\subsection{Performance of the \texorpdfstring{$\pi^w$}{} circuit}

    \begin{table}[!htb]
        \centering
        \scriptsize
        \setlength{\aboverulesep}{0pt}
        \setlength{\belowrulesep}{0pt}
        \renewcommand{\arraystretch}{1.5}
        \resizebox{\columnwidth}{!}{
        \begin{tabular}{|r|r|r||R{1.25cm}|R{1.25cm}|R{1.4cm}|R{1.6cm}|R{1.25cm}|R{1.6cm}|R{1.25cm}|R{1.25cm}|}\toprule
        
        \textbf{$k$} & \textbf{$n$} & \textbf{$(k+1)\,\cdot\,n$} & \textbf{\# constraints} & \textbf{key gen duration (s)} & \textbf{proof gen duration (s)} & \textbf{verification duration (s)} & \textbf{circuit size (MB)} & \textbf{verification key size (kB)} & \textbf{proof size (bytes)} \\
        \midrule
        4 & 20 & 100 & 26,894 & 84 & 2.981 & 0.238 & 15 & 25 & 707 \\
        4 & 30 & 150 & 34,340 & 93 & 3.619 & 0.239 & 20 & 25 & 710 \\
        4 & 40 & 200 & 42,834 & 113 & 4.530 & 0.245 & 23 & 25 & 707 \\
        4 & 50 & 250 & 50,272 & 117 & 5.091 & 0.239 & 26 & 25 & 705 \\
        4 & 100 & 500 & 89,819 & 177 & 8.641 & 0.236 & 46 & 25 & 708 \\
        4 & 150 & 750 & 128,362 & 213 & 12.080 & 0.238 & 62 & 25 & 706 \\
        4 & 200 & 1,000 & 168,619 & 294 & 15.996 & 0.241 & 86 & 25 & 706 \\
        4 & 250 & 1,250 & 208,162 & 337 & 19.410 & 0.237 & 102 & 25 & 709 \\
        4 & 300 & 1,500 & 246,658 & 372 & 22.400 & 0.241 & 118 & 25 & 706 \\
        4 & 350 & 1,750 & 286,670 & 508 & 27.322 & 0.241 & 150 & 25 & 707 \\
        4 & 400 & 2,000 & 325,975 & 558 & 31.217 & 0.241 & 166 & 25 & 707 \\
        4 & 450 & 2,250 & 364,709 & 627 & 33.409 & 0.238 & 182 & 25 & 707 \\
        4 & 500 & 2,500 & 404,962 & 659 & 36.868 & 0.237 & 198 & 25 & 708 \\
        4 & 550 & 2,750 & 444,030 & 713 & 40.447 & 0.242 & 241 & 25 & 708 \\
        4 & 600 & 3,000 & 482,522 & 803 & 43.499 & 0.246 & 230 & 25 & 707 \\
        4 & 650 & 3,250 & 522,775 & 990 & 49.498 & 0.242 & 246 & 25 & 707 \\
        4 & 700 & 3,500 & 562,317 & 1070 & 54.096 & 0.236 & 294 & 25 & 704 \\
        4 & 750 & 3,750 & 601,520 & 1089 & 57.042 & 0.237 & 310 & 25 & 705 \\
        4 & 800 & 4,000 & 640,825 & 1148 & 59.630 & 0.236 & 326 & 25 & 710 \\
        4 & 900 & 4,500 & 718,864 & 1246 & 70.394 & 0.234 & 358 & 25 & 706 \\
        4 & 1,000 & 5,000 & 798,659 & 1360 & 74.945 & 0.232 & 391 & 25 & 705 \\
        4 & 1,500 & 7,500 & 1,190,620 & 2571 & 120.736 & 0.263 & 614 & 25 & 705 \\
        \midrule
        
        \end{tabular}
        }
        \caption{Evaluation of essential parameters for building and verifying the $\pi^w$ circuit depending on $k$ and $n$.}
        \label{tab:evaluation_pi-w_full}
    \end{table}

\pagebreak
\subsection{Performance of the \texorpdfstring{$\pi^c$}{} circuit}

    \begin{table}[!htb]
        \centering
        \scriptsize
        \setlength{\aboverulesep}{0pt}
        \setlength{\belowrulesep}{0pt}
        \renewcommand{\arraystretch}{1.5}
        \resizebox{\columnwidth}{!}{
        \begin{tabular}{|r|r|r|r||R{1.25cm}|R{1.25cm}|R{1.4cm}|R{1.6cm}|R{1.25cm}|R{1.6cm}|R{1.25cm}|R{1.25cm}|}\toprule
        
        \textbf{$k$} & \textbf{$n$} & \textbf{$n_{\mathrm{test}}$} & \parbox[c]{1.4cm}{\textbf{$(k+1)\,\cdot\, (n+n_{\mathrm{test}})$}} & \textbf{\# constraints} & \textbf{key gen duration (s)} & \textbf{proof gen duration (s)} & \textbf{verification duration (s)} & \textbf{circuit size (MB)} & \textbf{verification key size (kB)} & \textbf{proof size (bytes)} \\
        \midrule
        4 & 20 & 10 & 150 & 20,959 & 82 & 3.005 & 0.243 & 11 & 4.2 & 707 \\
        4 & 30 & 10 & 200 & 26,940 & 90 & 3.472 & 0.237 & 13 & 4.2 & 706 \\
        4 & 40 & 10 & 250 & 33,971 & 108 & 4.439 & 0.241 & 18 & 4.2 & 704 \\
        4 & 50 & 10 & 300 & 39,952 & 114 & 4.951 & 0.250 & 21 & 4.2 & 710 \\
        4 & 100 & 10 & 550 & 72,194 & 174 & 8.152 & 0.244 & 38 & 4.2 & 705 \\
        4 & 150 & 15 & 825 & 105,940 & 206 & 11.108 & 0.243 & 51 & 4.2 & 708 \\
        4 & 200 & 20 & 1,100 & 141,684 & 293 & 15.439 & 0.243 & 74 & 4.2 & 707 \\
        4 & 250 & 25 & 1,375 & 176,240 & 339 & 18.568 & 0.247 & 88 & 4.2 & 704 \\
        4 & 500 & 50 & 2,750 & 349,881 & 659 & 36.949 & 0.244 & 174 & 4.2 & 708 \\
        4 & 1,000 & 100 & 5,500 & 697,265 & 1380 & 73.696 & 0.243 & 346 & 4.2 & 705 \\
        \midrule
        \end{tabular}
        }
        \caption{Evaluation of essential parameters for building and verifying the $\pi^c$ circuit depending on $k$, $n$, and $n_{\mathrm{test}}$.}
        \label{tab:evaluation_pi-c}
    \end{table}

\end{document}